\documentclass{article}
\usepackage{amssymb,amsfonts,amsthm,graphicx,psfig}
\def\pmb#1{\setbox0=\hbox{$#1$}%
  \kern-.025em\copy0\kern-\wd0
  \kern.05em\copy0\kern-\wd0
  \kern-.025em\raise.0433em\box0}
\def\pmbs#1{\setbox0=\hbox{$\scriptstyle #1$}%
  \kern-.0175em\copy0\kern-\wd0
  \kern.035em\copy0\kern-\wd0
  \kern-.0175em\raise.0303em\box0}
\def\be{\begin{equation}}
\def\ee{\end{equation}}
\def\bea{\begin{eqnarray}}
\def\eea{\end{eqnarray}}
\def\lb{\label}
\def\ct{\cite}
\def\bi{\bibitem}
\def\vec#1{\mbox{\boldmath$#1$}}
\def\ten#1{\mbox{\boldmath$#1$}}

\def\lgth{[\,\mbox{length}\,]}

\def\gam{\gamma}
\def\d{\delta}
\def\eps{\epsilon}
\def\Th{\Theta}
\def\sig{\sigma}

\def\Sig{\Sigma}

\def\om{\omega}
\def\Om{\Omega}
\def\Oml{\Omega_{\Lambda}}
\def\udot{\dot{u}}
\def\Udot{\dot{U}}
\def\cv{{\cal V}}

\def\cn{{\cal N}}

\def\cK{{\cal K}}
\def\cT{{\cal T}}
\def\ck{\Omega_{k}}

\newcommand{\cA}{\mathcal{A}}
\newcommand{\cC}{\mathcal{C}}

\def\vece{\vec{e}}
\def\vecu{\vec{u}}

\def\ptl{\partial}
\def\parb{\pmb{\partial}}
\def\e{{\rm e}}
\def\la{\langle}
\def\ra{\rangle}
\def\ti{\tilde}
\def\hsp5{\hspace{5mm}}
\newcommand{\sfrac}[2]{{\textstyle{#1\over#2}}}
\def\case#1/#2{\textstyle\frac{#1}{#2}}

\def\apj{{Astrophys.~J.} }
\def\cmp{{Commun.~Math.~Phys.} }
\def\cqg{{Class.~Quantum~Grav.} }
\def\grg{{Gen.~Rel.~Grav.} }
\def\jmp{{J.~Math.~Phys.} }
\def\mn{{Mon.~Not.~R.~Astron.~Soc.} }
\def\prd{{Phys.~Rev.~D} }
\def\prl{{Phys.~Rev.~Lett.} }
\def\prs{{Proc.~R.~Soc.~Lond.~A} }
\newcommand{\enl}{\\\hfill\rule{0pt}{0pt}}
\begin{document}
\title{{\sc The past attractor in inhomogeneous cosmology}}
\author{{\sc Claes Uggla$^{1}$\thanks{Electronic address: {\tt
	Claes.Uggla@kau.se}}, \,
Henk van Elst$^{2}$\thanks{Electronic address: {\tt
    H.van.Elst@qmul.ac.uk}}}\ , \\ 
{\sc John Wainwright$^{3}$\thanks{Electronic address: {\tt
	jwainwri@math.uwaterloo.ca}} \ and
George F.~R.~Ellis$^{4,2}$\thanks{Electronic address: {\tt
	ellis@maths.uct.ac.za}}} \\
$^{1}${\small\em Department of Physics, University of Karlstad,
S-651 88 Karlstad, Sweden} \\
$^{2}${\small\em Astronomy Unit, Queen Mary, University of London,
Mile End Road, London E1 4NS, United Kingdom} \\
$^{3}${\small\em Department of Applied Mathematics, University of
Waterloo, Waterloo, Ontario, Canada N2L 3G1} \\
$^{4}${\small\em Cosmology Group, Department of Mathematics and
Applied Mathematics, University of Cape Town}\\
{\small\em Rondebosch~7701, South Africa}}

\date{\normalsize{April 1, 2003}}
\maketitle
\begin{abstract}
We present a general framework for analyzing spatially
inhomogeneous cosmological dynamics. It employs Hubble-normalized
scale-invariant variables which are defined within the orthonormal
frame formalism, and leads to the formulation of Einstein's field
equations with a perfect fluid matter source as an autonomous
system of evolution equations and constraints. This framework
incorporates spatially homogeneous dynamics in a natural way as a
special case, thereby placing earlier work on spatially homogeneous
cosmology in a broader context, and allows us to draw on experience
gained in that field using dynamical systems methods. One of our
goals is to provide a precise formulation of the approach to the
spacelike initial singularity in cosmological models, described
heuristically by Belinski\v{\i}, Khalatnikov and
Lifshitz. Specifically, we construct an invariant set which we
conjecture forms the local past attractor for the evolution
equations. We anticipate that this new formulation will provide the
basis for proving rigorous theorems concerning the asymptotic
behavior of spatially inhomogeneous cosmological models.

\end{abstract}
\begin{flushleft}
PACS number(s): 04.20.-q, 98.80.Jk, 04.20.Dw, 04.20.Ha \\
Preprint number(s): QMUL-AU-2003-001, uct-cosmology-03/01,
gr-qc/0304002
\end{flushleft}

\section{Introduction}
\lb{sec:intro}
Scales and scale invariance play a crucial r\^{o}le in practically
all branches of physics, and general relativity (GR) and cosmology
are no exceptions.\footnote{See, e.g., the recent Resource Letter
by Wiesenfeld on scale invariance in physics and
beyond~\ct{wie2001}.} In these cases one is interested in
self-gravitating systems, which in the cosmological context
requires a matter model as well as a spacetime description.  This
in turn requires consideration of scales. In modern cosmology one
assumes that (today) there exists a global scale --- that of the
particle horizon. The empirical data are usually interpreted as
follows: on sufficiently large spatial scales, say a few percent of
the particle horizon, everything looks statistically roughly the
same in all directions. Combined with the Copernican principle
(``we are not located at a special place''), this suggests that one
can replace a very complicated matter distribution by a simple one:
a smooth distribution that is spatially homogeneous and isotropic,
obtained by averaging over sufficiently large spatial scales. Then
it is further assumed that one can also approximate the geometry of
the spacetime by a spatially homogeneous and isotropic geometry,
i.e., one assumes that the geometrical features trace those of the
matter and that possible ``excited geometrical modes'', like
gravitational waves, are negligible on these scales. This then
leads to modeling the cosmological spacetime by a Robertson--Walker
(RW) geometry.

The assumption of a RW geometry subsequently forces the summed
matter content to take the form of a perfect fluid through
Einstein's field equations (EFE). Although the mathematically
simplest matter model is a single perfect fluid, we need more
complex matter models to describe the real Universe. Indeed, matter
in the Universe consists of many components: at least (i) radiation
(photons), (ii) baryonic matter, (iii) neutrinos, (iv) dark matter,
and (v) dark energy/quintessence (other components like
cosmological magnetic fields are usually neglected). Once the
matter content has been specified and equations of state, scalar
field potentials, particle distribution functions, etc., have been
chosen, the evolution of the model is determined by the EFE and the
matter equations, e.g., the evolution equation for a scalar
field. This then leads to a Friedmann--Lema\^{\i}tre (FL) model for
the Universe.

The next step, aimed at describing the actual inhomogeneous
Universe, is to perturb the FL model and describe the evolution of
large-scale structures in the Universe, which appear at many scales
--- filaments and voids, superclusters of galaxies, galaxies
etc. But it is generally believed that linear perturbation theory
can account for them all on these large scales. And, indeed, the FL
scenario and the linear perturbations thereof (``almost-FL
models'') form a remarkably successful framework --- it seems to
consistently account for present observational evidence, at least
over sufficiently large smoothing scales. Moreover, it forms an
interpretational framework that encourages and steers further
observations. These are currently focused on determining the
various density contributions $\Om_{i}$ (including $\Oml$ for the
cosmological constant), and the spectrum and growth of density
perturbations. This is the simplest scenario consistent with
current observations.

Nevertheless, there are issues that need elucidation that by
necessity lie outside the domain of the standard almost-FL
picture. Here are some of them:

\begin{itemize}

\item To investigate the constraints observations impose on the
spacetime geometry of the Universe requires investigating a
hierarchy of more general models, perhaps characterized by assumed
``priors'' (where removing a prior necessarily involves looking
beyond ones favorite model, with the hope of getting further
support for it!).

\item To understand how special the FL models are and put them into
a broader context requires looking beyond them.

\item While the Universe is well described by almost-FL models at
present, this may not always have been true, and it may not remain
true in the far future; in particular, we would like to know the
largest class of models that can look like a FL model at some stage
of their history.

\item Using a FL and a linearized FL scenario rules out possible
non-linear effects, but these may dominate; e.g., as structures
become much more dense while aggregating to smaller scales.

\item GR is a highly non-linear theory; to prove that the linear
theory is correct requires going beyond linear perturbations.

\item The averaging and fitting procedures motivating the FL models
do not {\em a priori\/} commute with the EFE, i.e., starting with
an inhomogeneous model and smoothing it does not necessarily lead
to the model that has been smoothed from the outset and then
perturbed. This gives rise to a number of questions, e.g., can
inhomogeneities affect the overall evolution?  How do they affect
observations?

\item There are deep connections between GR, cosmology and
thermodynamics, e.g., relating (gravitational) entropy and the
arrow of time. To better understand such connections requires a
state space picture describing the set of solutions, where one can
examine coarse-graining and existence of attractors on this state
space, toward which the evolving cosmological models move. Since
entropy requires counting of {\em possible\/} states this requires
looking at models beyond FL.

\item What is the detailed nature of possible singularities? A
better understanding of generic features of singularities and their
dependence on matter content and initial data might shed light on
how the real Universe evolved initially. There might also exist at
least a local mathematical connection between the initial
singularity and the singularities of gravitational collapse. To
understand such a relationship, or its non-existence, again
requires an inspection of cosmological solutions beyond the
restrictions imposed by RW geometries.

\item A better classical understanding of singularities might help
to produce gravitational theories with greater domains of validity;
e.g., finding asymptotic symmetries of the field equations when
approaching singularities may provide sufficient structure to
asymptotically quantize the theory in a regime where quantum
gravity is supposed to be of importance.

\end{itemize}

Thus, there is ample motivation to probe a larger subset of the
cosmological solution space of the EFE than just the almost-FL
models. Our first goal in this paper is to develop a framework for
this purpose.

In view of the above-mentioned importance of scale invariance in
physics, we propose to introduce scale-invariant variables, and to
describe the evolution of a cosmological model by an orbit in an
infinite-dimensional dynamical state space, governed by first-order
(in time) autonomous evolution equations derived from the EFE and
the matter equations. The behavior of the model in the asymptotic
regimes, i.e., near the initial singularity and at late times, can
then possibly be described by a past attractor and a future
attractor of the evolution equations.

The appropriate mathematical vehicle for implementing this proposal
is the orthonormal frame formalism, since (i) it describes the
essential degrees of freedom of the gravitational field in a
coordinate independent manner, (ii) orthonormal frame vectors
provide local reference scales and so it allows one to naturally
introduce scale-invariant variables, and (iii) it leads directly to
first-order (in time) autonomous evolution equations.

To be more specific, by a cosmological model we mean a
four-dimensional spacetime manifold ${\cal M}$ endowed with a
Lorentzian metric $\ten{g}$ which satisfies the EFE with an
appropriate matter/energy distribution. We assume the existence of
a local foliation of the spacetime manifold by a one-parameter
family of spacelike 3-surfaces with a future-directed unit normal
congruence $\vec{u}$. We naturally choose this unit vector field to
be the timelike vector field in the orthonormal frame. We assume
that the cosmological model is {\em expanding\/}, i.e., the volume
expansion rate\footnote{Here $\Th = - ({\rm tr}\, k)$, where $\Th$
is the volume expansion rate of the normal congruence and $({\rm
tr}\,k)$ is the trace of the extrinsic curvature of the spacelike
3-surfaces.} $\Th$ of the normal congruence is positive. Because we
are working in a cosmological setting we will replace $\Th$ by the
Hubble scalar\footnote{When evaluated at the present epoch, the
Hubble scalar equals the Hubble constant $H_{0}$, familiar from
observational cosmology.} $H = \sfrac{1}{3}\,\Th$.

In the orthonormal frame formalism, the frame vector components and
the commutation functions are the basic dynamical variables for
describing gravitational fields, and they each have physical
dimension\footnote{We will use units such that Newton's
gravitational constant $G$ and the speed of light in vacuum $c$ are
given by $8\pi G/c^{2} = 1$ and $c = 1$.}  $\lgth^{-1}$.  The
Hubble scalar also has physical dimension $\lgth^{-1}$, and
constitutes the natural cosmological length scale through the
Hubble radius $H^{-1}$. This fact motivates one of the key steps in
our approach, namely, the introduction of {\em Hubble-normalized
variables\/} by dividing the frame vector components and the
commutation functions by $H$. Curvature quantities such as the
matter density and the orthonormal frame components of the Weyl
curvature tensor have physical dimension $\lgth^{-2}$, and hence
are normalized by dividing by $H^{2}$. This process of
Hubble-normalization has two important consequences. Firstly,
dimensional variables are replaced by dimensionless ones, leaving
$H$ as the only variable carrying physical dimensions. Secondly,
one is essentially factoring out the overall expansion of the
Universe, thereby measuring the dynamical importance of physical
quantities (e.g., the matter density) relative to the overall
expansion (cf.~Kristian and Sachs~\ct{krisac66}). This choice also
provides a link between mathematical analysis and observation,
since key observational variables are Hubble-normalized. Earlier
investigations of the asymptotic dynamics of cosmological models
using scale-invariant variables dealt with {\em spatially
homogeneous (SH) cosmologies\/}, i.e., models that admit a
three-parameter group of isometries acting on spacelike 3-surfaces
(see Wainwright and Ellis (WE) \ct{waiell97} and other references
therein), and the so-called {\em $G_{2}$ cosmologies\/}, which
admit a two-parameter group of isometries acting on spacelike
2-surfaces (see Refs.~\ct{hveetal2002} and \ct{hewwai90}). The
framework that we develop in this paper generalizes a program
that has been extremely successful in a SH context to a completely
general spatially inhomogeneous setting, i.e., to models that admit
no isometries. Indeed, the SH and $G_{2}$ cosmologies will be
incorporated in a natural way as invariant sets of the
infinite-dimensional Hubble-normalized state space. For brevity,
and to emphasize that they admit no isometries, we shall refer to
the models under consideration as {\em $G_{0}$ cosmologies\/}.

The framework that we are developing can be used to study both
asymptotic regimes in an ever-expanding cosmological model. In this
paper we focus on the initial singularity. The problem of
asymptotics in spacetimes that exhibit no symmetries poses a
formidable challenge. Nevertheless, concerning the existence of
singularities, remarkable progress was made several decades ago by
Penrose and Hawking, leading to their singularity
theorems~\ct{hawpen70,hawk73}. However, the singularity theorems do
not tell us much about the nature of the singularities. Detailed
asymptotic analysis, using the full EFE, is required for this
purpose. To date rigorous results have, with few exceptions, been
confined to cosmological models with isometries, in particular SH
and $G_{2}$ cosmologies. We shall discuss these results in
Sec.~\ref{sec:cossym}. As regards initial singularities in $G_{0}$
cosmologies, heuristic results were obtained by Belinski\v{\i},
Khalatnikov and Lifshitz (BKL)~\ct{bkl70,bkl82} by making {\em ad
hoc\/} metric assumptions that were subsequently inserted into the
EFE with the purpose of showing that they were consistent.  This
analysis led to a remarkable, although heuristic, conjecture that
has become part of the folklore of relativistic cosmology.

\bigskip
\noindent
{\em The BKL conjecture\/}:

For almost all cosmological solutions of Einstein's field
equations, a spacelike initial singularity is {\em
vacuum-dominated, local and oscillatory\/}. \enl

For cosmological models with a perfect fluid matter source, the
phrase ``vacuum-dominated'', or, equivalently, ``matter is not
dynamically significant'', is taken to mean that the
Hubble-normalized matter density (i.e., the density parameter
$\Om$) tends to zero at the initial singularity. The phrase ``for
almost all'' is needed because there are a number of exceptional
cases. Firstly, if the perfect fluid has a stiff equation of state,
the density parameter does not tend to zero (see Andersson and
Rendall~\ct{andren2001}). Secondly, there is a special type of
initial singularity called an {\em isotropic initial
singularity\/}, in the neighborhood of which the solution is
approximated locally by a spatially flat FL model (see Goode and
Wainwright~\ct{goowai85}), with the result that the density
parameter tends to the value~1. Isotropic initial singularities,
however, only arise from initial data that form a set of measure
zero.\footnote{There is in fact a wide variety of known special SH
and $G_{2}$ solutions in which the initial singularity is
matter-dominated, i.e., $\Om$ does not tend to zero. Like the
isotropic initial singularities, these singularities only arise
from initial data that form a set of measure zero.}

The word ``local'' in the BKL conjecture means heuristically that
the evolution at different spatial points effectively decouples as
the initial singularity is approached, with the result that
geometrical information propagation is asymptotically
eliminated. It is natural to describe this phenomenon as {\em
asymptotic silence of the gravitational field dynamics\/}. We shall
refer to the associated initial singularity as being a {\em silent
initial singularity\/}.\footnote{For further discussions on the
suppression of information propagation and asymptotic silence, see
Ref.~\ct{hveetal2002}.}

The word ``oscillatory'' in the BKL conjecture means that the
evolution into the past along a typical timeline passes through an
infinite sequence of Kasner states, generalizing the behavior
first encountered in the so-called Mixmaster models (SH models of
Bianchi Type--IX; see Misner~\ct{mis69}).

Our second main goal in this paper is to give a precise statement
of the BKL conjecture, within the framework of the
Hubble-normalized state space.

The plan of this paper is as follows. In Sec.~\ref{sec:eqs} we derive
the Hubble-normalized evolution equations and constraints for
$G_{0}$ cosmologies that arise from the EFE and the matter
equations. In Sec.~\ref{sec:state} we make a choice of gauge and then
describe some features of the Hubble-normalized state space, in
particular the SH invariant set and the silent boundary. We then
define the notion of a silent initial singularity. In
Sec.~\ref{sec:att}, by analyzing the dynamics on the silent boundary,
we are led to construct an invariant set which we conjecture is the
local past attractor for $G_{0}$ cosmologies with a silent initial
singularity, thereby making precise the notion of an oscillatory
singularity. In Sec.~\ref{sec:cossym} we consider various classes of
cosmological models with isometries and use the past attractor to
predict whether the initial singularity is oscillatory or not. We
conclude in Sec.~\ref{sec:disc} with a discussion of silent initial
singularities and the BKL conjecture, and raise some issues for
future study. Useful mathematical relations, such as the
propagation laws for the constraints and expressions for the
Hubble-normalized components of the Weyl curvature tensor, have
been gathered in an appendix.

\section{Evolution equations and constraints}
\lb{sec:eqs}
In this paper, we consider spatially inhomogeneous cosmological
models with a positive cosmological constant, $\Lambda$, and a
perfect fluid matter source with a linear barotropic equation of
state. We thus have
\be
\lb{eos}
\ti{p}(\ti{\mu}) = (\gam-1)\,\ti{\mu} \ ,
\ee
where $\ti{\mu}$ is the total energy density (assumed to be
non-negative) and $\ti{p}$ the isotropic pressure, in the rest
3-spaces associated with the fluid 4-velocity vector field
$\ti{\vecu}$, while $\gam$ is a constant parameter. The range
$$
1 \leq \gam \leq 2
$$
is of particular physical interest, since it ensures that the
perfect fluid satisfies the dominant and strong energy conditions
and the causality requirement that the speed of sound should be
less than or equal to that of light. The values $\gam = 1$ and
$\gam = \sfrac{4}{3}$ correspond to incoherent pressure-free matter
(``dust'') and incoherent radiation, respectively. In cosmology it
is natural to single out a future-directed timelike reference
congruence $\vece_{0} = \vec{u}$ of unit magnitude. This gives rise
to a (1+3)-decomposition of the perfect fluid
energy--momentum--stress tensor
\be
T_{ab} = \mu \, u_{a}\,u_{b} + 2\,q_{(a}\,u_{b)}
+ p\,h_{ab} + \pi_{ab} \ , 
\ee
with
\bea
\begin{array}{cclcccccl}
\lb{e3}
\mu & = & \Gamma^{2}\,G_{+}\,{\tilde \mu} & & & &
p & = & G_{+}^{-1}\,[\ (\gam-1)
+ (1-\sfrac{2}{3}\,\gam)\,v^{2}\ ]\,\mu \\ \\
q^{a} & = & \gam\,G_{+}^{-1}\,\mu v^{a} & & & &
\pi_{ab} & = & \gam\,G_{+}^{-1}\,\mu v_{\la a}v_{b\ra}
\end{array} \ .
\eea
The vector field $\vec{v}$, which represents the peculiar velocity
of the fluid relative to the rest 3-spaces of $\vece_{0}$, is
defined by
\be
\lb{jwe4}
\tilde{u}^{a} := \Gamma(u^{a} + v^{a}) \ , \hspace{10mm}
v_{a}u^{a} = 0 \ ,
\ee
with the Lorentz factor given by
\be
\lb{jwe5}
\Gamma := \frac{1}{\sqrt{1-v^{2}}} \ , \hspace{10mm}
v^{2} := v_{a} v^{a} \ .
\ee
The scalars $G_{\pm}$ (we shall require $G_{-}$ later) are defined
by
\be
\lb{jwe6}
G_{\pm} := 1 \pm (\gam-1)\,v^{2} \ .
\ee
To obtain an orthonormal frame, $\{\,\vece_{a}\,\}_{a = 0,1,2,3}$,
we supplement the timelike reference congruence $\vece_{0}$ with an
orthonormal spatial frame $\{\,\vece_{\alpha}\,\}_{\alpha = 1,2,3}$
in the rest 3-spaces of $\vece_{0}$. The frame metric is then given
by $\eta_{ab} = \mbox{diag}\,[\,-\,1, 1, 1, 1\,]$. In the
orthonormal frame formalism, introduced in relativistic cosmology,
among others, by Ellis~\ct{ell67}, the basic variables are the
frame vector components, the commutation functions associated with
the frame and the matter variables, and the dynamical equations are
provided by the EFE, the Jacobi identities and the contracted
Bianchi identities (the latter, for a perfect fluid, corresponding
to the relativistic extension of Euler's equations). We will make
use of an extended version of this formalism given by van Elst and
Uggla~\ct{hveugg97}. The dynamical equations consist of two sets,
those containing the temporal frame derivative $\vece_{0}$, which
we refer to as {\em evolution equations\/}, and those not
containing $\vece_{0}$, which we refer to as {\em constraints\/}.

To convert the dynamical equations of the orthonormal frame
formalism to a system of partial differential equations (PDE), it
is necessary to introduce a set of local coordinates
$\{x^{\mu}\}_{\mu = 0,1,2,3} = \{t,x^{i}\}_{i = 1,2,3}$. We do so
by adopting the standard (3+1)-approach (see, e.g.,
Refs.~\ct{yor79} and~\ct{wal84}). Here $\vece_{0}$ is assumed to be
{\em vorticity-free\/} and, thus, hypersurface-orthogonal. As is
well-known, this gives rise to a local foliation of the spacetime
manifold ${\cal M}$ by a one-parameter family of spacelike
3-surfaces, ${\cal S}$:$\{t=\mbox{constant}\}$. The (3+1)-approach
leads to the following coordinate expressions for the frame vector
fields (cf. Ref.~\ct{hveugg97}): \enl
\be
\lb{13cocom}
\vece_{0} = N^{-1}\,(\ptl_{t}-N^{i}\,\ptl_{i}) \ ,
\hspace{15mm}
\vece_{\alpha} = e_{\alpha}{}^{i}\,\ptl_{i} \ ,
\ee
where $N$ and $N^{i}$ are known as the lapse function and the shift
vector field, respectively.

\subsection{Dimensional equation system}
\lb{subsec:deqs}
We now present the dynamical equations as given in
Ref.~\ct{hveugg97}, simplified by the assumption that $\vece_{0}$
is vorticity-free $(\om^{\alpha} = 0)$. We begin with the
commutator equations, which serve to introduce the basic
gravitational field variables, and which will later be used to
derive some additional evolution equations and constraints.  \enl

\noindent
{\em Commutator equations\/}: \nopagebreak
\bea
\lb{comts}
\left[\,\vece_{0}, \vece_{\alpha}\,\right](f) & = &
\dot{u}_{\alpha}\,\vece_{0}(f)
- (H\,\d_{\alpha}{}^{\beta} + \sig_{\alpha}{}^{\beta}
- \eps_{\alpha\gam}{}^{\beta}\,\Om^{\gam})\,\vece_{\beta}(f) \\
\lb{comss}
0 & = & (C_{\rm com})_{\alpha\beta}(f)
\ := \ \left[\,\vece_{\alpha}, \vece_{\beta}\,\right](f)
- (2a_{[\alpha}\,\d_{\beta]}{}^{\gam}
+ \eps_{\alpha\beta\delta}\,n^{\delta\gam})\,\vece_{\gam}(f) \ ,
\eea
with $f$ denoting an arbitrary real-valued spacetime scalar. Here
$H$ is the Hubble scalar which is related to the volume expansion
rate $\Th$ of $\vece_{0}$ according to $H := \sfrac{1}{3}\,\Th$.
The quantities $\udot^{\alpha}$ and $\sig_{\alpha\beta}$ are the
acceleration and shear rate of $\vece_{0}$, respectively, while
$\Om^{\alpha}$ describes the angular velocity of the spatial frame
$\{\,\vece_{\alpha}\,\}$ along the integral curves of $\vece_{0}$
relative to a Fermi-propagated one. The quantities $a^{\alpha}$ and
$n_{\alpha\beta}$ determine the connection on the spacelike
3-surfaces ${\cal S}$:$\{t=\mbox{constant}\}$. \enl

\noindent
{\em Einstein's field equations, Jacobi identities and contracted
Bianchi identities (Euler's equations)\/}: \enl \nopagebreak

\noindent
{\em Evolution equations\/}:
\bea
\lb{hdot}
\vece_{0}(H)
& = & -\,H^{2}
- \sfrac{1}{3}\,(\sig_{\alpha\beta}\sig^{\alpha\beta})
- \sfrac{1}{6}\,(\mu+3p) + \sfrac{1}{3}\,\Lambda
+ \sfrac{1}{3}\,(\vece_{\alpha}
+ \udot_{\alpha}-2a_{\alpha})\,(\udot^{\alpha}) \\
\lb{adot}
\vece_{0}(a^{\alpha})
& = & -\,(H\,\d^{\alpha}{}_{\beta} + \sig^{\alpha}{}_{\beta}
- \eps^{\alpha}{}_{\gam\beta}\,\Om^{\gam})\,a^{\beta}
- \sfrac{1}{2}\,(\vece_{\beta}+\dot{u}_{\beta})\,
(2H\,\d^{\alpha\beta} - \sig^{\alpha\beta}
- \eps^{\alpha\beta}{}_{\gam}\,\Om^{\gam}) \\ 
\lb{sigdot}
\vece_{0}(\sig^{\alpha\beta})
& = & -\,3H\,\sig^{\alpha\beta} 
- 2n^{\la\alpha}{}_{\gam}\,n^{\beta\ra\gam}
+ n_{\gam}{}^{\gam}\,n^{\la\alpha\beta\ra}  
- \d^{\gam\la\alpha}\,\vece_{\gam}(a^{\beta\ra}) \nonumber \\
& & + \ \eps^{\gam\delta\la\alpha}\,
[\,(\vece_{\gam}+\udot_{\gam}-2a_{\gam})\,(n^{\beta\ra}{}_{\delta})
+ 2\Om_{\gam}\,\sig^{\beta\ra}{}_{\delta}\,]
+ \pi^{\alpha\beta} + (\d^{\gam\la\alpha}\,\vece_{\gam}
+\udot^{\la\alpha}+a^{\la\alpha})\,(\udot^{\beta\ra}) \\
\lb{ndot}
\vece_{0}(n^{\alpha\beta})
& = & -\,(H\,\d^{(\alpha}{}_{\delta}
- 2\sig^{(\alpha}{}_{\delta}
- 2\eps^{\gam}{}_{\delta}{}^{(\alpha}\,\Om_{\gam})\,n^{\beta)\delta}
- (\vece_{\gam}+\udot_{\gam})\,
(\eps^{\gam\delta(\alpha}\,\sig^{\beta)}{}_{\delta}
- \d^{\gam(\alpha}\,\Om^{\beta)} + \d^{\alpha\beta}\,\Om^{\gam}) \\
\lb{mudot}
\vece_{0}(\mu)
& = & -\,3H\,(\mu+p)
- (\vece_{\alpha}+2\udot_{\alpha}-2a_{\alpha})\,(q^{\alpha})
- (\sig_{\alpha\beta}\pi^{\alpha\beta}) \\
\lb{valpha}
\vece_{0}(v^{\alpha})
& = & \frac{G_{+}}{\gam\,G_{-}\,\mu}\,
\left[\ -\,\gam\,v^{\alpha}\,\vece_{0}(\mu)
+ (G_{-}\d^{\alpha}{}_{\beta}
+ 2(\gam-1)v^{\alpha}v_{\beta})\,\vece_{0}(q^{\beta})
\ \right] \ ,
\eea
where
\bea
\lb{qalpha}
\vece_{0}(q^{\alpha})
& = & -\,(4H\,\d^{\alpha}{}_{\beta} + \sig^{\alpha}{}_{\beta}
- \eps^{\alpha}{}_{\gam\beta}\,\Om^{\gam})\,q^{\beta}
- \d^{\alpha\beta}\,\vece_{\beta}(p)
- (\mu+p)\,\udot^{\alpha} \nonumber \\
& & \hsp5 - \ (\vece_{\beta}+\udot_{\beta}-3a_{\beta})\,
(\pi^{\alpha\beta}) + \eps^{\alpha\beta\gam}\,n_{\beta\delta}\,
\pi_{\gam}{}^{\delta} \ .
\eea

\noindent
{\em Constraints\/}:
\bea
\lb{gauss}
0 & = & (C_{\rm G}) \ := \ 2(2\vece_{\alpha}
-3a_{\alpha})\,(a^{\alpha}) - (n_{\alpha\beta}n^{\alpha\beta})
+ \sfrac{1}{2}\,(n_{\alpha}{}^{\alpha})^{2}
+ 6H^{2} - (\sig_{\alpha\beta}\sig^{\alpha\beta})
- 2\mu - 2\Lambda \\
\lb{codacci}
0 & = & (C_{\rm C})^{\alpha}
\ := \ -\,\vece_{\beta}(2H\,\d^{\alpha\beta}-\sig^{\alpha\beta})
- 3a_{\beta}\,\sig^{\alpha\beta}
- \eps^{\alpha\beta\gam}\,n_{\beta\delta}\,\sig_{\gam}{}^{\delta}
+ q^{\alpha} \\
\lb{jacobi1}
0 & = & (C_{\rm J})^{\alpha}
\ := \ \vece_{\beta}(n^{\alpha\beta}
+\eps^{\alpha\beta\gam}\,a_{\gam}) - 2a_{\beta}\,n^{\alpha\beta}
\ .
\eea 
Note that we are {\em not\/} provided with evolution equations for
any of the\footnote{Employing the terminology of
Friedrich~\ct{fri96}, Sec.~5.2.} coordinate gauge source functions
$N$ and $N^{i}$ (which reside in $\vece_{0}$) or the frame gauge
source functions $\udot^{\alpha}$ and $\Om^{\alpha}$. Note also
that these ten gauge source functions do {\em not\/} appear in the
constraints. Independent of a choice of gauge (to be discussed in
Sec.~\ref{sec:state}), the evolution equations (\ref{comts}) and
(\ref{hdot})--(\ref{qalpha}) propagate the constraints
(\ref{comss}) and (\ref{gauss})--(\ref{jacobi1}) along the integral
curves of $\vece_{0}$ according to
Eqs.~(\ref{ccomdot})--(\ref{cjac1dot}) in the appendix.

There are, in addition, two {\em gauge constraints\/} that restrict
four of the gauge source functions, given by
\bea
\lb{31coj}
0 & = & (C_{\omega})^{\alpha}
\ := \ [\,\eps^{\alpha\beta\gam}\,(\vece_{\beta}
- a_{\beta}) - n^{\alpha\gam}\,]\,\udot_{\gam} \\
\lb{jwe21}
0 & = & (C_{\dot{u}})_{\alpha}
\ := \ N^{-1}\,\vece_{\alpha}(N) - \udot_{\alpha} \ .
\eea
The former is a consequence of assuming $\vece_{0}$ to be
vorticity-free, the latter follows from Eq.~(\ref{comts}) upon
substitution of Eq.~(\ref{13cocom}). The propagation of the gauge
constraints along the integral curves of $\vece_{0}$ can be
established once a choice of temporal gauge has been made (as this
determines what the currently unknown frame derivatives
$\vece_{0}(\udot^{\alpha})$ and $\vece_{0}(N)$ should be).

\subsection{Scale-invariant equation system}
\lb{subsec:dleqs}
We now introduce Hubble-normalized frame, connection and matter
variables as follows:
\be
\lb{jw22}
\parb_{0} := \frac{1}{H}\,\vece_{0} \ , \hspace{10mm}
\parb_{\alpha} := \frac{1}{H}\,\vece_{\alpha} \ ,
\ee
\bea
\lb{dlcon}
\{\,\Udot^{\alpha},
\,\Sig_{\alpha\beta}, \,A^{\alpha},
\,N_{\alpha\beta}, \,R^{\alpha}\,\}
& := & \{\,\udot^{\alpha},
\,\sig_{\alpha\beta}, \,a^{\alpha},
\,n_{\alpha\beta}, \,\Om^{\alpha}\,\}/H \\
\lb{dlcurv}
\{\,\Om, \,\Oml, \,P, \,Q^{\alpha}, \,\Pi_{\alpha \beta} \}
& := & \{\,\mu, \,\Lambda , \,p, \,q^{\alpha},
\,\pi_{\alpha\beta}\}/(3H^{2}) \ .
\eea
It follows from Eq.~(\ref{e3}) that
\be
\lb{dlpqpi}
P = G_{+}^{-1}\,[\ (\gam-1)+
(1-\sfrac{2}{3}\,\gam)\,v^{2}\ ]\,\Om \ , \hspace{10mm}
Q^{\alpha} = \frac{\gam}{G_{+}}\,\Om v^{\alpha} \ , \hspace{10mm}
\Pi_{\alpha\beta} = \frac{\gam}{G_{+}}\,\Om v_{\la\alpha}
v_{\beta\ra} \ .
\ee
Expressing the Hubble-normalized frame derivatives $\parb_{0}$ and
$\parb_{\alpha}$ with respect to the local coordinates introduced
in Eq.~(\ref{13cocom}) leads to:
\be
\lb{dl13cocom}
\parb_{0} = \cn^{-1}\,(\ptl_{t}-N^{i}\,\ptl_{i}) \ , \hspace{10mm}
\parb_{\alpha} = E_{\alpha}{}^{i}\,\ptl_{i} \ ,
\ee
where
\be
\lb{e26}
\cn := NH \ , \hspace{10mm}
E_{\alpha}{}^{i} := \frac{e_{\alpha}{}^{i}}{H} \ .
\ee
In order to write the dimensional equation system in
Hubble-normalized form, it is necessary to introduce the {\em
deceleration parameter\/} $q$ and the {\em spatial Hubble
gradient\/} $r_{\alpha}$, defined by
\bea
\lb{jw27}
(q+1) & := &  -\,\frac{1}{H}\,\parb_{0}H \\
\lb{jw28}
r_{\alpha} & := & -\,\frac{1}{H}\,\parb_{\alpha}H \ .
\eea
The definition~(\ref{jw27}), together with Raychaudhuri's
equation~(\ref{hdot}) and Eqs.~(\ref{dlcon}) and~(\ref{dlcurv}), lead to
the following key expression for $q$:
\be
\lb{hdecel}
q = 2\Sig^{2}
+ \sfrac{1}{2}\,(\Om+3P) - \Oml
- \sfrac{1}{3}\,(\parb_{\alpha}
-r_{\alpha}+\Udot_{\alpha}-2A_{\alpha})\,\Udot^{\alpha} \ ,
\ee
where $\Sig^{2} :=
\sfrac{1}{6}\,(\Sig_{\alpha\beta}\Sig^{\alpha\beta})$.

We now use Eqs.~(\ref{jw27}) and (\ref{jw28}) to write the commutator
equations (\ref{comts}) and (\ref{comss}) in Hubble-normalized
form. The result is
\bea
\lb{E3}
\left[\,\parb_{0}, \parb_{\alpha}\,\right]f & = &
-\,(r_{\alpha}-\dot{U}_{\alpha})\,\parb_{0}f
+ (q\,\d_{\alpha}{}^{\beta} - \Sig_{\alpha}{}^{\beta}
+ \eps_{\alpha\gam}{}^{\beta}\,R^{\gam})\,\parb_{\beta}f \\
\lb{E4}
0 & = & ({\cal C}_{\rm com})_{\alpha\beta}(f)
\ := \ \left[\,\parb_{\alpha}, \parb_{\beta}\,\right]f
- \left[\ 2\,(r_{[\alpha}+A_{[\alpha})\,\d_{\beta]}{}^{\gam}
+ \eps_{\alpha\beta\delta}\,N^{\delta\gam}\ \right]
\,\parb_{\gam}f \ .
\eea

We now write the evolution equations (\ref{hdot})--(\ref{qalpha}) and
the constraints (\ref{gauss})--(\ref{jacobi1}) in Hubble-normalized
form.\footnote{In explicit component form these equations are
available online at the URL given in Ref.~\ct{hve2002}. An earlier
scale-invariant equation system (based on an orthonormal frame
formulation), which employs the once-contracted second Bianchi
identities and Weyl curvature variables, was derived by two of the
authors (H.v.E. and C.U.) and given in Ref.~\ct{hve96}, but no
specific choice of temporal gauge or spatial frame was introduced
then.} \enl

\noindent
{\em Evolution equations\/}:
\bea
\lb{dladot}
\parb_{0}A^{\alpha}
& = & (q\,\d^{\alpha}{}_{\beta} - \Sig^{\alpha}{}_{\beta}
+ \eps^{\alpha}{}_{\gam\beta}\,R^{\gam})\,A^{\beta}
- \sfrac{1}{2}\,(\parb_{\beta}-r_{\beta}+
\Udot_{\beta})\,(2\d^{\alpha\beta} - \Sig^{\alpha\beta}
-\eps^{\alpha\beta}{}_{\gam}\,R^{\gam}) \\
\lb{dlsigdot}
\parb_{0}\Sig^{\alpha\beta}
& = & (q-2)\,\Sig^{\alpha\beta} - 2N^{\la\alpha}{}_{\gam}\,
N^{\beta\ra\gam} + N_{\gam}{}^{\gam}\,N^{\la\alpha\beta\ra}
- \d^{\gam\la\alpha}\,(\parb_{\gam}-r_{\gam})\,A^{\beta\ra}
\nonumber \\
& & \hsp5 + \ \eps^{\gam\delta\la\alpha}\,[\,(\parb_{\gam}-r_{\gam}
+\Udot_{\gam}-2A_{\gam})\,N^{\beta\ra}{}_{\delta}
+ 2R_{\gam}\,\Sig^{\beta\ra}{}_{\delta}\,] + 3\Pi^{\alpha\beta}
\nonumber \\
& & \hsp5 + \ (\d^{\gam\la\alpha}\,\parb_{\gam}-r^{\la\alpha}
+\Udot^{\la\alpha}+A^{\la\alpha})\,\Udot^{\beta\ra} \\
\lb{dlndot}
\parb_{0}N^{\alpha\beta}
& = & (q\,\d^{(\alpha}{}_{\delta}
+ 2\Sig^{(\alpha}{}_{\delta}
+ 2\eps^{\gam}{}_{\delta}{}^{(\alpha}\,R_{\gam})\,N^{\beta)\delta}
\nonumber \\
& & \hsp5 - \ (\parb_{\gam}-r_{\gam}+\Udot_{\gam})\,
(\eps^{\gam\delta(\alpha}\,\Sig^{\beta)}{}_{\delta}
- \d^{\gam(\alpha}\,R^{\beta)} + \d^{\alpha\beta}\,R^{\gam}) \\
\lb{dlomdot}
\parb_{0}\Om
& = & (2q-1)\,\Om - 3P
- (\parb_{\alpha}-2r_{\alpha}+2\Udot_{\alpha}-2A_{\alpha})\,
Q^{\alpha} - (\Sig_{\alpha\beta}\Pi^{\alpha\beta}) \\
\lb{dlvdot}
\parb_{0}v^{\alpha}
& = & \frac{G_{+}}{\gam\,G_{-}\,\Om}\,
\left[\ -\,\gam\,v^{\alpha}\,(\parb_{0}-2q-2)\,\Om
+ (G_{-}\d^{\alpha}{}_{\beta}
+ 2(\gam-1)v^{\alpha}v_{\beta})\,(\parb_{0}-2q-2)\,Q^{\beta}
\ \right] \\
\lb{dlomldot}
\parb_{0}\Oml
& = & 2\,(q+1)\,\Oml \ ,
\eea
where\footnote{We give the Hubble-normalized relativistic Euler
equations, Eqs.~(\ref{dlomdot}) and (\ref{dlvdot}), in explicit form in
the appendix; see Eqs.~(\ref{dlomdotf}) and (\ref{dlvdotf}).}
\bea
\lb{dlqalpha}
\parb_{0}Q^{\alpha}
& = & [\,2\,(q-1)\,\d^{\alpha}{}_{\beta} - \Sig^{\alpha}{}_{\beta}
+ \eps^{\alpha}{}_{\gam\beta}\,R^{\gam}\,]\,Q^{\beta}
- \d^{\alpha\beta}\,(\parb_{\beta}-2r_{\beta})\,P
- (\Om+P)\,\Udot^{\alpha} \nonumber \\
& & \hsp5 - \ (\parb_{\beta}-2r_{\beta}+\Udot_{\beta}
-3A_{\beta})\,\Pi^{\alpha\beta} + \eps^{\alpha\beta\gam}\,
N_{\beta\delta}\,\Pi_{\gam}{}^{\delta} \ .
\eea

\noindent
{\em Constraints\/}:
\bea
\lb{dlgauss}
0 & = & ({\cal C}_{\rm G})
\ := \ 1 - \ck - \Sig^{2} - \Om - \Oml \\
\lb{dlcodacci}
0 & = & ({\cal C}_{\rm C})^{\alpha}
\ := \ \parb_{\beta}\Sig^{\alpha\beta}
+ (2\d^{\alpha}{}_{\beta}-\Sig^{\alpha}{}_{\beta})\,r^{\beta}
- 3A_{\beta}\,\Sig^{\alpha\beta}
- \eps^{\alpha\beta\gam}\,N_{\beta\delta}\,\Sig_{\gam}{}^{\delta}
+ 3Q^{\alpha} \\
\lb{dljacobi1}
0 & = & ({\cal C}_{\rm J})^{\alpha}
\ := \ (\parb_{\beta}-r_{\beta})\,(N^{\alpha\beta}
+\eps^{\alpha\beta\gam}\,A_{\gam}) - 2A_{\beta}\,N^{\alpha\beta}
\\
\lb{dloml}
0 & = & ({\cal C}_{\Lambda})_{\alpha}
\ := \ (\parb_{\alpha} - 2r_{\alpha})\,\Oml \ ,
\eea
where
\be
\lb{omkdef}
\ck := -\,\sfrac{1}{3}\,(2\parb_{\alpha}
-2r_{\alpha}-3A_{\alpha})\,A^{\alpha}
+ \sfrac{1}{6}\,(N_{\alpha\beta}N^{\alpha\beta})
- \sfrac{1}{12}\,(N_{\alpha}{}^{\alpha})^{2} \ .
\ee
We have also included an evolution equation and a constraint for
$\Omega_{\Lambda}$, which are a direct consequence of
Eqs.~(\ref{dlcurv}), (\ref{jw27}) and (\ref{jw28}).

The r\^{o}le of the spatial Hubble gradient $r_{\alpha}$ requires
comment. One can use the Codacci constraint~(\ref{dlcodacci}) to
express $r_{\alpha}$ in terms of Hubble-normalized variables.  The
resulting formula for $r_{\alpha}$ involves the inverse of the
matrix $(2\d^{\alpha}{}_{\beta}-\Sig^{\alpha}{}_{\beta})$, and in
order to avoid this algebraic complication, we propose to treat
$r_{\alpha}$ as a dependent variable. Choosing $f = H$ in the
commutator equations~(\ref{E3}) and (\ref{E4}), and making use of
Eqs.~(\ref{jw27}) and (\ref{jw28}), leads to both an evolution equation
and a constraint for $r_{\alpha}$:
\bea
\lb{dlrdot}
\parb_{0}r_{\alpha}
& = & (q\,\d_{\alpha}{}^{\beta} - \Sig_{\alpha}{}^{\beta}
+ \eps_{\alpha\gam}{}^{\beta}\,R^{\gam})\,r_{\beta}
+ (\parb_{\alpha}-r_{\alpha}+\Udot_{\alpha})\,(q+1) \\
\lb{dlrcon}
0 & = & ({\cal C}_{r})^{\alpha}  
\ := \ [\,\eps^{\alpha\beta\gam}\,(\parb_{\beta} - A_{\beta})
- N^{\alpha\gam}\,]\,r_{\gam} \ .
\eea
These equations constitute integrability conditions for
Eqs.~(\ref{jw27}) and (\ref{jw28}).

When we write the evolution equations and constraints as PDE by
expressing $\parb_{0}$ and $\parb_{\alpha}$ in terms of partial
derivatives using Eq.~(\ref{dl13cocom}), the frame components
$E_{\alpha}{}^{i}$ enter into the equations as dependent
variables. Successively choosing $f = x^{i}$, $i = 1,2,3$, in the
commutator equations~(\ref{E3}) and (\ref{E4}) leads to an evolution
equation and a constraint for $E_{\alpha}{}^{i}$:
\bea
\lb{dl13comts}
\parb_{0}E_{\alpha}{}^{i}
& = & (q\,\d_{\alpha}{}^{\beta} - \Sig_{\alpha}{}^{\beta}
+ \eps_{\alpha\gam}{}^{\beta}\,R^{\gam})\,E_{\beta}{}^{i}
- \cn^{-1}\,\parb_{\alpha}N^{i} \\
\lb{dl13comss}
0 & = & ({\cal C}_{{\rm com}})^{i}{}_{\alpha\beta}
\ := \ 2\,(\parb_{[\alpha}-r_{[\alpha}-A_{[\alpha})\,
E_{\beta]}{}^{i}
- \eps_{\alpha\beta\delta}\,N^{\delta\gam}\,E_{\gam}{}^{i} \ .
\eea
Finally, we give the Hubble-normalized form of the gauge
constraints~(\ref{31coj}) and (\ref{jwe21}):
\bea
\lb{dl31coj}
0 & = & ({\cal C}_{W})^{\alpha}
\ := \ [\,\eps^{\alpha\beta\gam}\,(\parb_{\beta} - r_{\beta}
- A_{\beta}) - N^{\alpha\gam}\,]\,\Udot_{\gam} \\
\lb{dl13comtt31}
0 & = & ({\cal C}_{\Udot})_{\alpha}
\ := \ \parb_{\alpha}\ln\cn
+ (r_{\alpha}-\Udot_{\alpha}) \ .
\eea
%

\section{Gauge fixing and the Hubble-normalized state space}
\lb{sec:state}
In the previous section we presented a constrained system of
coupled PDE that govern the evolution of $G_{0}$ cosmologies. The
dependent variables are

\begin{itemize}

\item the spatial frame vector field components $E_{\alpha}{}^{i}$,

\item the spatial Hubble gradient $r_{\alpha}$, and

\item the gravitational field and matter variables
$\Sig_{\alpha\beta}$, $A^{\alpha}$, $N_{\alpha\beta}$, $\Om$,
$v^{\alpha}$, and $\Oml$.

\end{itemize}

\noindent
The system of PDE is underdetermined due to the presence of the
gauge source functions
$$
\cn, \ N^{i}, \ \Udot^{\alpha}, \ R^{\alpha} \ ,
$$
which reflects the fact that there is freedom in the choice of the
local coordinates and of the orthonormal frame. We now use this
gauge freedom to specify the gauge source functions, and then
proceed to describe some aspects of the Hubble-normalized state
space.

\subsection{Fixing the gauge}
\lb{subsec:gaugefix}
We begin by using the coordinate freedom to set the shift vector
field in Eqs.~(\ref{13cocom}) and (\ref{dl13cocom}) to zero:
\be
\lb{zeroshift}
N^{i} = 0 \ .
\ee
We then choose the timelike reference congruence $\vece_{0}$ so
that
\be
\lb{cn0}
\parb_{\alpha}\cn = 0 \ .
\ee
We are then free to specialize the time coordinate $t$ so that
\be
\lb{N1}
\cn = 1 \ .
\ee
The effect of these choices is that the lapse function in
Eq.~(\ref{13cocom}) is given by $N = H^{-1}$, as follows from
Eq.~(\ref{e26}). The gauge constraint (\ref{dl13comtt31}), taken in
conjunction with the above conditions, reduces to
\be
\lb{sepudot}
0 = ({\cal C}_{\Udot})_{\alpha}^{\rm sv}
= (r_{\alpha}-\Udot_{\alpha})
\hsp5 \Rightarrow \hsp5
\Udot_{\alpha} = r_{\alpha} \ ,
\ee
thus determining the frame gauge source functions
$\Udot_{\alpha}$. The advantage of making the
choices~(\ref{zeroshift}) and (\ref{N1}) is that the temporal frame
derivative $\parb_{0}$, given by Eqs.~(\ref{dl13cocom}), simplifies
to a partial derivative,
\be
\parb_{0} = \ptl_{t} \ .
\ee
The combined gauge choices~(\ref{zeroshift}) and~(\ref{N1}) have a
simple geometrical interpretation in terms of the {\em volume
density\/} $\cv$ associated with the family of spacelike 3-surfaces
${\cal S}$:$\{t=\mbox{constant}\}$, which is defined by
\be
\cv^{-1} := \det(e_{\alpha}{}^{i})\ .
\ee
Using Eq.~(\ref{zeroshift}), the commutator equations yield
\be
\lb{dlvolder}
\cn^{-1}\,\frac{\ptl_t\cv}{\cv} = 3 \ , \hspace{15mm}
E_{\alpha}{}^{i}\,\frac{\ptl_{i}\cv}{\cv} = -\,2A_{\alpha}
- \ptl_{i}E_{\alpha}{}^{i} + r_{\alpha} \ .
\ee
It follows with Eq.~(\ref{N1}) that
\be
\lb{sepvol}
\cv = \ell_{0}^{3}\,\e^{3 t}\,\hat{m} \ ,
\ee
where $\hat{m} = {\hat m}(x^i)$ is a freely specifiable positive
real-valued function of $x^{i}$, which we consider given, and
$\ell_{0}$ is the unit of the physical dimension $\lgth$.  We thus
refer to this gauge choice as the {\em separable volume
gauge\/}. Note that the reduced gauge constraint~(\ref{sepudot})
propagates along $\vece_{0}$ according to Eq.~(\ref{gfcpropsa}) in
the appendix. {\em This ensures the local existence (in time) of
the separable volume gauge.\/}

Equations~(\ref{dlvolder}) and (\ref{sepvol}) subsequently
yield the constraint
\be
\lb{dlsepcon}
0 = ({\cal C}_{A})_{\alpha} 
= A_{\alpha} + \sfrac{1}{2}\left(\,\ptl_{i}E_{\alpha}{}^{i}
- r_{\alpha} + E_{\alpha}{}^{i}\,\ptl_{i}\ln\hat{m}\,\right) \ .
\ee
Finally we use a time- and space-dependent rotation of the spatial
frame to relate the frame gauge source functions $R^{\alpha}$ to
the off-diagonal components of the shear rate tensor according
to\footnote{In contrast to the present frame choice, one can use
the frame freedom to reduce the number of variables, e.g., by
diagonalizing the shear rate tensor. However, the present choice
leads to great simplification of the equations when it comes to
analyzing the past attractor. There are other useful choices; in
particular, when one has a preferred spatial direction induced by
an isometry. In such a case it is often advantageous to choose the
$R^{\alpha}$-component associated with the preferred direction to
have the opposite sign compared with the present choice.}
\be
\lb{frame}
(R_{1},R_{2},R_{3})^{T} = (\Sig_{23},\Sig_{31},\Sig_{12})^{T} \ .
\ee
At this stage there is no freedom remaining in the choice of
frame.\footnote{With the exception of the special cases when the
shear rate tensor is locally rotationally symmetrical or zero; when
the frame is uniquely determined, all the Hubble-normalized
variables employed are scalar invariants.} The coordinate freedom
is
$$
t' = t + \mbox{constant} \ , \hspace{10mm}
x^{i}{}^{\prime} = f^{i}(x^{j}) \ .
$$

An important question, which we do not pursue at present, except in
a footnote in Sec.~\ref{sec:cossym}, is to what extent the analysis
in this paper (in particular the construction of the past
attractor) depends on the choice of temporal gauge. Here we use the
separable volume gauge, as defined by Eqs.~(\ref{zeroshift}) and
(\ref{N1}), which appears to be particularly well-adapted to
Hubble-normalized variables. For $G_{2}$ cosmologies, which we
shall refer to later, the usual and most convenient temporal gauge
is the so-called separable area gauge (see, e.g.,
Ref.~\ct{hveetal2002}).

\subsection{Hubble-normalized state space}
\lb{subsec:state}
\subsubsection{Overview}
The Hubble-normalized state vector for $G_{0}$ cosmologies is given
by
\be
\lb{stateX}
\vec{X} = (E_{\alpha}{}^{i}, r_{\alpha}, \Sig_{\alpha\beta}, 
N_{\alpha\beta}, A^{\alpha},\Om, v^{\alpha}, \Oml)^{T} \ .
\ee
The evolution equations and constraints in the previous section can
be written concisely in the form
\bea
\lb{evolX}
\ptl_{t}\vec{X} & = & 
\vec{F}(\vec{X}, \ptl_{i}\vec{X}, \ptl_{i}\ptl_{j}\vec{X}) \\
\lb{constrX}
0 & = & \vec{C}(\vec{X}, \ptl_{i}\vec{X}) \ ,
\eea
with the spatial derivatives appearing linearly (apart from the
evolution equation for $r_{\alpha}$). A surprising feature of the
evolution equations is that they contain second-order spatial
derivatives; in this respect they are reminiscent of a system of
quasi-linear diffusion equations. The only second-order spatial
derivatives in the evolution equations, however, are those of the
spatial Hubble gradient, $\ptl_{i}\ptl_{j}r_{\alpha}$, and they
appear only in the evolution equation for $r_{\alpha}$ itself. They
arise due to the fact that in the separable volume gauge the
deceleration parameter $q$ contains the first spatial derivatives
of $r_{\alpha}$. In fact, in the separable volume gauge
Eq.~(\ref{hdecel}) for $q$ assumes the form
\be
\lb{qsep}
q = 2\Sig^{2}
+ G_{+}^{-1} \left[\,(3\gam-2) + (2-\gam)v^{2}\,\right] \Om
- \Oml - \sfrac{1}{3}\,(\parb_{\alpha}-2A_{\alpha})\,r^{\alpha} \ .
\ee
The term $\parb_{\alpha}q$ in the evolution equation for
$r_{\alpha}$, Eq.~(\ref{dlrdot}), thus contains
$\ptl_{i}\ptl_{j}r_{\alpha}$.

A second noteworthy feature of the system of PDE~(\ref{evolX}) is
that the evolution equation for $E_{\alpha}{}^{i}$ is {\em
homogeneous\/}, which implies that the equation $E_{\alpha}{}^{i} =
0$ defines an invariant set. We shall discuss the significance of
this set later in this section.  In order to clearly exhibit these
aspects of the evolution equations, we now decompose the
Hubble-normalized state vector~(\ref{stateX}) as follows
\be
\lb{XY}
\vec{X} = (E_{\alpha}{}^{i}, r_{\alpha})^{T} \oplus \vec{Y} \ ,
\ee
where
\be
\lb{Y}
\vec{Y} =
(\Sig_{\alpha\beta}, N_{\alpha\beta}, A^{\alpha},
\Om, v^{\alpha}, \Oml)^{T} \ .
\ee 
We can now write the system (\ref{evolX}) in a more explicit form
as follows:
\be
\lb{Esepvol}
\ptl_{t}E_{\alpha}{}^{i}
= (q\,\d_{\alpha}{}^{\beta} - \Sig_{\alpha}{}^{\beta}
+ \eps_{\alpha\gam}{}^{\beta}\,R^{\gam})\,E_{\beta}{}^{i}\ ,
\ee
with $q$ given by Eq.~(\ref{qsep}), and
\bea
\lb{rsep}
\ptl_{t}r_{\alpha} & = &
[\,G_{\alpha}{}^{\beta}(\vec{Y})
+ \sfrac{2}{3}\,(\parb_{\alpha}+r_{\alpha})A^{\beta}\,]\,r_{\beta}
+ G_{\alpha}{}^{\beta\gam}(\vec{Y})\,\parb_{\beta}r_{\gam}
- \sfrac{1}{3}\,\parb_{\alpha}(\parb_{\beta}r^{\beta})
+ \parb_{\alpha}G(\vec{Y}) \\
\lb{evolY}
\ptl_{t}Y_{A} & = & F_{A}(\vec{Y})
+ F_{A}{}^{B\alpha}(\vec{Y})\,\parb_{\alpha}Y_{B}
+ F_{A}{}^{\alpha\beta}(\vec{Y})\,\parb_{\alpha}r_{\beta}
+ F_{A}{}^{\alpha}(\vec{Y})\,r_{\alpha}\ ,
\eea
with
$$
\parb_{\alpha} = E_{\alpha}{}^{i}\,\ptl_{i} \ .
$$
The coefficients $G_{\alpha}{}^{\beta}$,
$G_{\alpha}{}^{\beta\gam}$, $G$, $F_{A}$, $F_{A}{}^{B\alpha}$,
$F_{A}{}^{\alpha\beta}$ and $F_{A}{}^{\alpha}$ are functions of the
components of $\vec{Y}$.

\subsubsection{Spatially homogeneous cosmologies}
\lb{subsec:shcosm}
We now discuss how the SH cosmologies are described within the
$G_{0}$ framework. These are obtained by requiring that the spatial
frame derivatives of the gravitational field and matter variables
$\vec{Y}$, and of the normalization factor $H$, be zero, i.e.,
\be
\lb{8}
\parb_{\alpha}\vec{Y} = 0 \ , \hspace{10mm}
r_{\alpha} = 0 \ .
\ee
It then follows that all the dimensional commutation functions and
matter variables are constant on the spacelike 3-surfaces ${\cal
S}$:$\{t=\mbox{constant}\}$, which are thus the orbits of a
three-parameter group of isometries. The evolution equations
(\ref{rsep}) and (\ref{evolY}) imply that the SH restrictions (\ref{8})
define an invariant set of the full evolution equations, which we
shall call the {\em SH invariant set\/}. Indeed, Eq.~(\ref{rsep})
is trivially satisfied, and Eq.~(\ref{evolY}) reduces to a system
of ordinary differential equations, namely
\be
\lb{888}
\ptl_{t}Y_{A} = F_{A}(\vec{Y}) \ .
\ee
The non-trivial constraints defined by $(\cC_{\rm G})$, $(\cC_{\rm
C})^{\alpha}$ and $(\cC_{\rm J})^{\alpha}$
[\,cf.~Eqs.~(\ref{dlgauss})--(\ref{dljacobi1})\,] become purely
algebraical restrictions on $\vec{Y}$, which we write symbolically
as
\be
\lb{SHconY}
\cC(\vec{Y}) = 0 \ .
\ee
An important aspect of this process of specialization is that the
evolution equation (\ref{Esepvol}) for $E_{\alpha}{}^{i}$ decouples
from the evolution equation for $\vec{Y}$, which means that {\em
the dynamics of SH cosmologies can be analyzed using only
Eqs.~(\ref{888}) and (\ref{SHconY})\/} (cf.~WE). In this context, one
can think of the variables $\vec{Y}$ as defining a {\em reduced
Hubble-normalized state space\/}, of finite dimension, for the SH
cosmologies.

In the SH context the restriction $v^{\alpha} = 0$ defines an
invariant subset, giving the so-called non-tilted SH cosmologies,
and the Bianchi classification of the isometry group leads to a
hierarchy of invariant subsets, some of which have been analyzed in
detail in the literature. For example, the conditions
\be
v^{\alpha} = 0 \ , \hsp5 A^{\alpha} = 0 \ ,\hsp5
N_{\alpha\beta} = 0 \ (\alpha\neq\beta) \ , \hsp5
R^{\alpha} = 0 \ , \hsp5
\Oml = 0 \ ,
\ee
give the non-tilted SH perfect fluid cosmologies of class A in the
canonical frame (see WE, Chap.~6, but with some differences in
notation).

Specializing further, by requiring the shear rate to be zero,
\be
\lb{88}
\Sig_{\alpha \beta} = 0 \ ,
\ee
in addition to conditions~(\ref{8}), we obtain the {\em FL invariant
set\,}, which describes the familiar Friedmann--Lema\^{\i}tre
cosmologies.  Equations~(\ref{8}) and (\ref{88}) imply that $v^{\alpha}
= 0$ and ${\cal S}_{\alpha\beta} = 0$, where ${\cal
S}_{\alpha\beta}$ is the tracefree part of the 3-Ricci curvature
(see App.~\ref{subsec:siweyl}), and hence that the spacelike
3-surfaces ${\cal S}$:$\{t=\mbox{constant}\}$ are of constant
curvature. In addition, the electric and magnetic parts of the Weyl
curvature (see App.~\ref{subsec:siweyl}) are zero, $0 = {\cal
E}_{\alpha\beta} = {\cal H}_{\alpha \beta}$. The deceleration
parameter simplifies to
$$
q = \sfrac{1}{2}\,(3\gam-2)\,\Om - \Oml \ .
$$
%

\subsubsection{Silent boundary}
\lb{subsec:silbound}
We noted earlier that, because the evolution
equation~(\ref{Esepvol}) for $E_{\alpha}{}^{i}$ is homogeneous, the
equation
\be
E_{\alpha}{}^{i} = 0
\ee
defines an invariant set of the full evolution
equations.\footnote{Note that this does not necessarily imply
$\lim_{t\rightarrow -\infty} E_{\alpha}{}^{i} = 0$.} In the
introduction we discussed the notion of a {\em silent initial
singularity\/}, which was introduced heuristically as an initial
singularity with the property that the evolution along neighboring
timelines decouples as the singularity is approached. In
Sec.~\ref{sec:att} we shall make a formal definition of a silent
initial singularity, but for now we note that a key requirement for
an initial singularity to be silent is
\be
\lb{assil}
\lim_{t\rightarrow -\infty} E_{\alpha}{}^{i} = 0 \ ,
\ee
i.e., the orbit that describes the evolution of the model is past
asymptotic to the invariant set $E_{\alpha}{}^{i} = 0$. We will
thus refer to this invariant set as the {\em silent boundary\/}.

On the silent boundary, the evolution equation (\ref{rsep}) for
$r_{\alpha}$ simplifies to the homogeneous form
\be
\lb{rhom}
\ptl_{t} r_{\alpha} = 
[\,G_{\alpha}{}^{\beta}(\vec{Y})
+ \sfrac{2}{3}\,r_{\alpha}\,A^{\beta}\,]\,r_{\beta} \ .
\ee
It follows that the equation
\be
r_{\alpha} = 0
\ee
defines an invariant subset of the silent boundary. On this
invariant subset the remaining evolution equation~(\ref{evolY})
reduces to
\be
\lb{Yhom}
\ptl_{t}Y_{A} = F_{A}(\vec{Y}) \ ,
\ee
which coincides with the evolution equation~(\ref{888}) for the SH
cosmologies. The remaining constraints are purely algebraical, and
can be written symbolically as
\be
\lb{e81}
\cC(\vec{Y}) = 0 \ . 
\ee
One thus obtains a representation of the SH dynamics on the
invariant set
\be
\lb{Er}
E_{\alpha}{}^{i} = 0 \ ,\hspace{10mm} r_{\alpha} = 0 \ ,
\ee
i.e., within the silent boundary. Since $E_{\alpha}{}^{i} = 0$,
however, the spatial dependence of the Hubble-normalized variables
$\vec{Y}$ is completely unrestricted, and hence these solutions of
the evolution equations and constraints do {\em not\/} in general
correspond to exact solutions of the EFE.

\section{Silent singularities and the generalized Mixmaster
  attractor}
\lb{sec:att}
In this section, we formalize the notion of a silent initial
singularity, which was introduced heuristically in
Sec.~\ref{sec:intro}.\footnote{The concepts we propose for
classifying an initial singularity as ``silent'' can be applied
analogously to final singularities.} We then construct an invariant
set in the silent boundary that we conjecture is the local past
attractor for $G_{0}$ cosmologies with a silent initial
singularity. The detailed structure of the past attractor in turn
relies heavily on our knowledge of the asymptotic dynamics near the
initial singularity in SH cosmologies.

\subsection{Silent initial singularities}
\lb{subsec:silinising}
In terms of Hubble-normalized variables and the separable volume
gauge, the spacelike initial singularity in a $G_{0}$ cosmology is
approached as $t \rightarrow -\infty$.  We now define a {\em silent
initial singularity\/} to be one which satisfies
\be
\lb{e82}
\lim_{t \rightarrow -\infty} E_{\alpha}{}^{i} = 0 \ ,
\ee
\be
\lb{e82a}
\lim_{t \rightarrow -\infty} r_{\alpha} = 0 \ ,
\ee
and
\be
\lb{e83}
\lim_{t \rightarrow -\infty} \parb_{\alpha}\vec{Y} = 0 \ ,
\ee
where the $E_{\alpha}{}^{i}$ are the Hubble-normalized components
of the spatial frame vectors [\,see Eq.~(\ref{dl13cocom})\,],
$r_{\alpha}$ is the spatial Hubble gradient [\,see
Eq.~(\ref{jw28})\,] and $\vec{Y}$ represents the Hubble-normalized
gravitational field and matter variables [\,see
Eq.~(\ref{Y})\,]. More precisely, we require that
Eqs.~(\ref{e82})--(\ref{e83}) are satisfied {\em along typical
timelines of $\vece_{0}$\/}.

One might initially think that the condition (\ref{e83}) is a
consequence of Eq.~(\ref{e82}), since
$$
\parb_{\alpha}\vec{Y} = E_{\alpha}{}^{i}\,
\frac{\ptl\vec{Y}}{\ptl x^{i}} \ .
$$
However, the analysis of Gowdy solutions with so-called spikes (see
Refs.~\ct{bermon93}, \ct{garwea2002} and \ct{renwea2001}) shows
that the partial derivatives $\ptl\vec{Y}/\ptl x^{i}$ can diverge
as $t \rightarrow -\infty$. Thus, the requirement~(\ref{e83})
demands that the $E_{\alpha}{}^{i}$ tend to zero sufficiently fast.

We now present some evidence to justify proposing the above
definition.  Firstly, for SH cosmologies, which we have seen
satisfy the restrictions~(\ref{8}), the evolution equation for the
$E_{\alpha}{}^{i}$ decouples from the equation for $\vec{Y}$.  This
evolution equation, in conjunction with the known results about the
asymptotic behavior of the variables $\vec{Y}$ (see WE, Chaps.~5
and~6, and Ringstr\"{o}m~\ct{rin2001}), provides strong evidence
that typical solutions satisfy\footnote{We are indebted to Hans
Ringstr\"{o}m for helpful discussions on this matter.} the
remaining requirement~(\ref{e82}) for a silent initial singularity.
An example of an exceptional class of SH solutions, i.e., solutions
for which the initial singularity is not silent, are those that are
past asymptotic to the flat Kasner solution (the Taub form of
Minkowski spacetime), given by
$$
{\rm d}s^{2} = -\,{\rm d}T^{2} + T^{2}\,{\rm d}x^{2}
+ \ell_{0}^{2}\,({\rm d}y^{2} + {\rm d}z^{2}) \ ,
$$
where $T$ is clock time. In terms of the dimensionless separable
volume time $t = \ln(T/\ell_{0})$, this line element reads
$$
\ell_{0}^{-2}\,{\rm d}s^{2} = e^{2t}\,(-\,{\rm d}t^{2}
+ {\rm d}x^{2}) + {\rm d}y^{2} + {\rm d}z^{2} \ .
$$
from which it follows that
$$
\lim_{t \rightarrow -\infty} E_{\alpha}{}^{i}
= {\rm diag}\,(3,0,0) \ .
$$

Secondly, further evidence is provided by recent research on
$G_{2}$ cosmologies, although the situation is clouded by the fact
that an area time gauge rather than the separable volume gauge is
used (but see the footnote in the next section about the gauge
issue). Indeed, one can use the asymptotic analysis of vacuum
orthogonally transitive $G_{2}$ cosmologies (the so-called Gowdy
solutions~\ct{gow71,gow74}; in the present context see in
particular Ref.~\ct{kicren98}) to show that the
conditions~(\ref{e82}) and~(\ref{e83}) are satisfied along typical
timelines, even when spikes occur. However, in general $G_{2}$
cosmologies the situation is more complicated and further studies
are needed to establish if condition~(\ref{e83}) holds or if it is
violated along exceptional timelines due to the presence of
spikes.\footnote{Woei Chet Lim, private communication.}

These results suggest that the notion of a silent initial
singularity may be of importance as regards the description of
generic spacelike initial singularities. Further support is
provided by heuristic arguments of a physical nature, as
follows. We anticipate that generic spacelike initial singularities
are associated with increasingly strong gravitational fields,
gradually approaching local curvature radii of Planck-scale order,
which will lead to the formation of {\em particle horizons} (see,
e.g., Rindler~\ct{rin56}). The existence of particle horizons is
governed by null geodesics, which satisfy
\be
\lb{E1}
1 = \d_{\alpha\beta}\left(E^{\alpha}{}_{i}\,
\frac{{\rm d}x^{i}}{{\rm d}t}\right)\left(E^{\beta}{}_{j}\,
\frac{{\rm d}x^{j}}{{\rm d}t}\right) \ ,
\ee
(see Eq. (\ref{ds2}) in the appendix), where $E^{\alpha}{}_{i}$ are
the components of the Hubble-normalized 1-forms associated with the
orthonormal frame:
\be
\lb{E2}
E^{\alpha}{}_{i}\,E_{\beta}{}^{i} = \d^{\alpha}{}_{\beta} \ .
\ee
If particle horizons form, we expect that the past-directed null
geodesics emanating from a chosen point $P$ will satisfy $x^{i}(t)
\rightarrow x_{H}^{i}$ (constants) and ${\rm d}x^{i}/{\rm d}t
\rightarrow 0$, as $t \rightarrow -\infty$. It follows from
Eq.~(\ref{E1}) that
$$
E^{\alpha}{}_{i}\,\frac{{\rm d}x^{i}}{{\rm d}t} \rightarrow
b^{\alpha} \ ,
$$
with $\d_{\alpha\beta}b^{\alpha}b^{\beta} = 1$, and, hence, that
$$
\left(\lim_{t \rightarrow -\infty}E_{\alpha}{}^{i}\right)
b^{\alpha} = 0 \ .
$$
Since this must hold for all null geodesics emanating from $P$,
$b^{\alpha}$ is arbitrary, implying that the limit~(\ref{e82}) holds.
In other words, we expect that the increasingly strong
gravitational fields associated with a typical spacelike initial
singularity will lead to the first condition in the proposed
definition of a silent initial singularity.

We now show heuristically that condition~(\ref{e83}) restricts the
scale of spatial inhomogeneities as the initial singularity is
approached. If $E_{\alpha}{}^{i}$ tends to zero at an exponentially
bounded rate as $t \rightarrow -\infty$ (as in SH and $G_{2}$
cosmologies), the coordinate distance to the particle horizon in a
direction $b^{\alpha}$ will also tend to zero at an exponentially
bounded rate:
$$
\Delta x_{H}^{i} \approx b^{\alpha}\,E_{\alpha}{}^{i}
\hsp5 {\rm as} \hsp5
t \rightarrow -\infty \ .
$$
The change $\Delta\vec{Y}$ in the Hubble-normalized variables
$\vec{Y}$ corresponding to a change $\Delta x_{H}^{i}$ is
approximated by
$$
\Delta\vec{Y} \approx \frac{\ptl\vec{Y}}{\ptl x^{i}}\,
\Delta x_{H}^{i}
\approx b^{\alpha}\,\parb_{\alpha}\vec{Y} \ .
$$
It thus follows from the limit~(\ref{e83}) that $\Delta \vec{Y}
\rightarrow 0$ as $t \rightarrow -\infty$. In other words, {\em the
physical significance of the limit~(\ref{e83}) is that spatial
inhomogeneities have super-horizon scale asymptotically as\/} $t
\rightarrow -\infty$, and, hence, up to the particle horizon scale
a solution is asymptotically SH.

With the preceding discussion as motivation we now make our first
conjecture. \enl

\hangindent=9em
{\bf Conjecture 1}: For almost all cosmological solutions of
Einstein's field equations, a spacelike initial singularity is
silent. \enl

\noindent
Proving this conjecture entails establishing the
limits~(\ref{e82})--(\ref{e83}).

\subsection{Stable set into the past}
\lb{subsec:stabset}
We think of the evolution of the Hubble-normalized state vector
$\vec{X}(t, x^{i})$, {\em for fixed\/} $x^{i}$, as being described
by an orbit in a {\em finite-dimensional\/} Hubble-normalized state
space. As $t \rightarrow -\infty$, this orbit will be asymptotic to
a {\em past attractor\/}, which, in accordance with the definition
of a silent initial singularity [\,see
Eqs.~(\ref{e82})--(\ref{e83})\,], will be contained in the subset of
the silent boundary defined by
\be
\lb{e84}
E_{\alpha}{}^{i} =0 \ ,
\hspace{10mm}
r_{\alpha} = 0 \ .
\ee
The evolution of a spatially inhomogeneous model is described by
infinitely many such orbits, each of which is asymptotic to the
past attractor. The details of the approach to the past attractor,
however, will depend on spatial position $x^{i}$, thereby
reflecting the spatial inhomogeneity of the model. On the other
hand, the evolution of an SH model will be described by a single
orbit. The essential point is that {\em the dynamics in the
invariant set\/} (\ref{e84}), {\em which govern the asymptotic
dynamics of both classes of models, is determined by the SH
evolution equations and constraints\/}, as shown in
Subsec.~\ref{subsec:silbound}.

The next step in constructing the putative past attractor is to
partition the Hubble-normalized state vector $\vec{X}$ into stable
and unstable variables, as regards evolution into the past.
Firstly, within our framework, the BKL conjecture means that
\be
\lb{e85}
\lim_{t \rightarrow -\infty}\Om = 0 \ ,
\hspace{10mm}
\lim_{t \rightarrow -\infty} \Oml = 0 \ ,
\ee
(i.e., the initial singularity is vacuum-dominated). Secondly,
asymptotic analysis and numerical experiments for SH cosmologies
and $G_{2}$ cosmologies suggest that
\be
\lb{e7}
\lim_{t \rightarrow -\infty} A^{\alpha} = 0 \ ,
\hspace{10mm}
\lim_{t \rightarrow -\infty} N_{\alpha\beta} = 0
\ (\alpha \neq \beta) \ ,
\ee
along a typical orbit. It is thus convenient to decompose the
Hubble-normalized state vector $\vec{X}$ as follows:
$$
\vec{X} = \vec{X}_{\rm s} \oplus \vec{X}_{\rm u} \ ,
$$
where
\be
\lb{e8}
\vec{X}_{\rm s} = \left( E_{\alpha}{}^{i},
r_{\alpha}, A^{\alpha}, N_{\alpha\beta}\,(\alpha \neq \beta), \Om,
\Oml\right)^{T} \ , 
\ee
and
\be
\lb{e9}
\vec{X}_{\rm u} = (\Sig_{\alpha}, R^{\alpha}, N_{\alpha},
v^{\alpha})^{T} \ .
\ee
Here, for brevity, we have written\footnote{Not to be confused with
the notation used in Ref.~\ct{hewetal2002}, where $\Sig_{1}$ was
defined to be equal to $\Sig_{23}$, and cycle on $(1, 2, 3)$.}
$$
\Sig_{\alpha} := \Sig_{\alpha \alpha} \ ,
\hspace{10mm}
N_{\alpha} := N_{\alpha\alpha} \ .
$$
In terms of this notation, our conjectures~(\ref{e82}), (\ref{e82a}),
(\ref{e85}) and (\ref{e7}) can be written
\be
\lb{e10}
\lim_{t \rightarrow - \infty} \vec{X}_{\rm s} = \vec{0} \ .
\ee
We shall refer to the variables $\vec{X}_{\rm s}$ as the {\em
stable variables\/}, and the remaining variables $\vec{X}_{\rm u}$
in Eq.~(\ref{e9}) as the {\em unstable variables\/}. We shall
provide evidence that the variables $\vec{X}_{\rm u}$ remain
bounded as $t \rightarrow -\infty$, but that their limits do not
exist. We note in passing that further justification for the
terminology ``stable'' and ``unstable'' in this context will be
provided shortly, when we show that the variables in $\vec{X}_{\rm
s}$ are stable on the Kasner circles, while the variables in
$\vec{X}_{\rm u}$ are unstable.

We now list the evolution equations on the subset $\vec{X}_{\rm s}
= 0$. Firstly, the variables $\Sig_{\alpha}$, $R^{\alpha}$ and
$N_{\alpha}$ satisfy
\bea
\lb{e16}
\ptl_{t}\Sig_{1} & = & 2(1-\Sig^{2})\,\Sig_{1}
+ 2(R_{2}^{2}-R_{3}^{2}) - 3S_{1} \\
\lb{e15}
\ptl_{t}R_{1} & = & \left[\,-2(1-\Sig^{2}) + \Sig_{2} - \Sig_{3}\,
\right] R_{1} \\
\lb{e14}
\ptl_{t}N_{1} & = & 2(\Sig^{2} + \Sig_{1})\,N_{1} \ ,
\eea
where
\be
\lb{e17}
S_{1} := \sfrac{2}{9}\,N_{1}^{2} - \sfrac{1}{3}\,N_{1}\,
(N_{2}+N_{3}) - \sfrac{1}{9}\,(N_{2} - N_{3})^{2} \ ,
\ee
and cycle on $(1,2,3)$. These variables are restricted by the
Gau\ss\ constraint~(\ref{dlgauss}), which now reads
\be
\lb{e18}
1 = \Sig^{2} + \sfrac{1}{6}\,(N_{1}^{2}+N_{2}^{2}+N_{3}^{2})
- \sfrac{1}{12}\,(N_{1}+N_{2}+N_{3})^{2} \ , 
\ee
with
\be
\lb{e19}
\Sig^{2} = \sfrac{1}{6}\,(\Sig_{1}^{2}+\Sig_{2}^{2}+\Sig_{3}^{2}
+2R_{1}^{2}+2R_{2}^{2}+2R_{3}^{2}) \ . 
\ee
Secondly, the evolution equation for $v^{\alpha}$ now reads
\be
\lb{e20}
\ptl_{t}v^{\alpha} = \frac{1}{G_{-}}\left[\,(3\gam-4)\,
(1-v^{2}) + (2-\gam)(\,\Sig_{\beta\gam}v^{\beta}v^{\gam})\,\right]
v^{\alpha}
- \left[\,\Sig^{\alpha}{}_{\beta} - \eps^{\alpha}{}_{\gam\beta}\,
(R^{\gam}+N^{\gam}{}_{\delta}v^{\delta})\,\right] v^{\beta} \ ,
\ee
where it is convenient to retain the index
notation. We note for future use that Eq.~(\ref{e20}) implies
\be
\lb{e21}
\ptl_{t}v^{2} = \frac{2}{G_{-}}\,(1 - v^{2})
\left[\,(3\gam-4)\,v^{2} - (\Sig_{\alpha\beta}v^{\alpha}
v^{\beta})\,\right] \ . 
\ee

Although the variables $\vec{X}_{\rm u}$ are unstable into the
past, it turns out that certain combinations of these unstable
variables are in fact stable. Firstly, the limit~(\ref{e10}), in
conjunction with the equation for $\ptl_{t}N_{\alpha\beta}\,(\alpha
\neq \beta)$ and the Codacci constraint, leads to the following
limits:
\be
\lb{e11}
\lim_{t \rightarrow - \infty} R_{\alpha} N_{\beta} = 0 \ ,
\hspace{10mm} \alpha \neq \beta \ .
\ee
As a result, the subset of the Hubble-normalized state space
defined by
\be
\lb{e12}
\vec{X}_{\rm s} = \vec{0}
\ee
is an invariant set only if the following restrictions hold:
\be
\lb{e13}
R_{\alpha} N_{\beta} = 0 \ , \hspace{10mm} \alpha \neq \beta \ .
\ee
The essential point is that the products $R_{\alpha}N_{\beta}$
$(\alpha \neq \beta)$ are stable into the past.

Secondly, we can make use of known results about SH models to
motivate another limit, in addition to Eq.~(\ref{e11}). We
introduce the function
\be
\lb{deltaN}
\Delta_{N} := (N_{1}N_{2})^{2} + (N_{2}N_{3})^{2}
+ (N_{3}N_{1})^{2} \ .
\ee
If $\Delta_{N} \neq 0$, i.e., if more than one $N_{\alpha}$ is
non-zero, then Eq.~(\ref{e13}) implies $R^{\alpha} = 0$. Then the
evolution equations (\ref{e16})--(\ref{e14}) reduce to the evolution
equations for vacuum SH models of class A. It has been
shown\footnote{See Ringstr\"{o}m~\ct{rin2001} for the case where
the $N_{\alpha}$ have the same sign (Bianchi Type--IX
case). Numerical simulations suggest that this result is also true
in the Bianchi Type--VIII case.} that solutions of these evolution
equations satisfy
\be
\lb{delta0}
\lim_{t \rightarrow -\infty} \Delta_{N} = 0 \ .
\ee
It is thus plausible that if Eqs.~(\ref{e10}) and (\ref{e11}) hold,
then so does Eq.~(\ref{delta0}). We shall refer to the invariant
set defined by
\be
\lb{stable}
\vec{X}_{\rm s} = \vec{0} \ , \hspace{10mm}
R_{\alpha}N_{\beta} = 0 \ (\alpha \neq \beta) \ , \hspace{10mm}
\Delta_{N} = 0 \ ,
\ee 
as the {\em stable subset\/} into the past and make the following
conjecture. \enl

\hangindent=9em
{\bf Conjecture 2}: The local past attractor $\mathcal{A}^{-}$ for
$G_{0}$ cosmologies with a silent initial singularity is a subset
of the stable subset. \enl

\noindent
Proving this conjecture entails proving the limits~(\ref{e10}),
(\ref{e11}) and (\ref{delta0}), assuming the validity of
(\ref{e82})--(\ref{e83}).

We believe that this conjecture can be strengthened, however. In
order to do this, we need to describe how the Kasner vacuum
solutions are represented within the present framework.

\subsection{Kasner circles}
The line element for the Kasner vacuum solutions is
$$
\ell_{0}^{-2}\,{\rm d}s^{2} = -\,{\rm d}T^{2} + T^{2p_{1}}\,{\rm d}x^{2}
+ T^{2p_{2}}\,{\rm d}y^{2} + T^{2p_{3}}\,{\rm d}z^{2} \ ,
$$
where the Kasner exponents $p_{1}$, $p_{2}$ and $p_{3}$ are
constants that satisfy
$$
p_{1} + p_{2} + p_{3} = 1 \ ,
\hspace{10mm}
p_{1}^{2} + p_{2}^{2} + p_{3}^{2} = 1 \ ,
$$
and $\ell_{0}T$ is clock time. The Kasner exponents can take values
that are described by the inequalities $-\sfrac{1}{3} \leq p_{1}
\leq 0 \leq p_{2} \leq \sfrac{2}{3} \leq p_{3} \leq 1$ (or
permutations thereof); see Ref.~\ct{lifkha63}, p.~196. Relative to
the natural orthonormal frame associated with this line element,
the Hubble-normalized connection variables are all zero except for
the shear rate tensor, which is diagonal and given by
$$
\Sig_{\alpha\beta} = {\rm diag}\,(3p_{1}-1, 3p_{2}-1, 3p_{3}-1) \ .
$$
One can also represent the Kasner solutions relative to a spatial
frame that is not Fermi-propagated, as is the spatial frame
specified by Eq.~(\ref{frame}). Some of these alternative
representations are important in what follows.

Within our formulation, all possible representations of the Kasner
solutions are given by
\be
\lb{e22}
0 = A^{\alpha} = N_{\alpha\beta} = \Om = v^{\alpha} = \Oml \ ,
\ee
\be
\lb{e23}
r_{\alpha} = 0 \ , \hspace{10mm}
\parb_{\alpha}\Sig_{\beta \gam} = 0 \ , \hspace{10mm}
\parb_{[\alpha}E_{\beta]}{}^{i} = 0 \ ,
\ee
with the $R^{\alpha}$ given by Eq.~(\ref{frame}). The Gau\ss\
constraint~(\ref{dlgauss}), together with Eqs.~(\ref{e18})
and~(\ref{qsep}), implies that
\be
\lb{e24}
\Sig^{2} = 1\ , \hspace{10mm}
q = 2 \ .
\ee
The evolution of the non-zero variables $E_{\alpha}{}^{i}$ and
$\Sig_{\alpha\beta}$ is governed by
\bea
\lb{e26a}
\ptl_{t}E_{\alpha}{}^{i}
& = & (2\d_{\alpha}{}^{\beta}-\Sig_{\alpha}{}^{\beta}
+ \eps_{\alpha\gam}{}^{\beta}\,R^{\gam})\,E_{\beta}{}^{i} \\
\lb{e25}
\ptl_{t}\Sig^{\alpha\beta}
& = & 2\eps^{\gam\delta\la\alpha}\,R_{\gam}\,
\Sig^{\beta\ra}{}_{\delta} \ ,
\eea
as follows from Eqs.~(\ref{Esepvol}) and~(\ref{dlsigdot}).

In the physical region of the Hubble-normalized state space, i.e.,
$\det (E_{\alpha}{}^{i}) \neq 0$, Eqs.~(\ref{e23}) imply that
$\Sig_{\alpha\beta} = \Sig_{\alpha\beta}(t)$, and that the spatial
coordinate freedom can be used to obtain $E_{\alpha}{}^{i} =
E_{\alpha}{}^{i} (t)$, confirming that the Kasner solutions are SH
and of Bianchi Type--I. On the silent boundary $(E_{\alpha}{}^{i} =
0)$, however, Eqs.~(\ref{e23}) become trivial, with the result that
{\em the spatial dependence of $\Sig_{\alpha\beta}$ is
unrestricted\/}. One thus obtains a representation of the Kasner
dynamics locally on the silent boundary, even though the line
element, as given by Eq.~(\ref{ds2}) in the appendix, is singular.
Indeed, the Kasner dynamics on the silent boundary is described by
the orbits that satisfy $\vec{X}_{\rm s} = \vec{0}$ and the
additional restriction $N_{\alpha} = 0$, as follows from
Eqs.~(\ref{e8}) and (\ref{e22}). We shall refer to this subset,
defined by
\be
\lb{e27}
\vec{X}_{\rm s} = \vec{0} \ , \hspace{10mm} N_{\alpha} = 0 \ , 
\ee
as the {\em Kasner set on the silent boundary\/}.

The evolution equations on the Kasner set are obtained by setting
$\Sig^{2} =1$ and $S_{\alpha} = 0$ in Eqs.~(\ref{e16})
and~(\ref{e15}), which yields:
\bea
\lb{e29}
\ptl_{t}\Sig_{1} & = & 2(R_{2}^{2}-R_{3}^{2})\\
\lb{e28}
\ptl_{t} R_{1} & = & (\Sig_{2}-\Sig_{3})\,R_{1} \ ,
\eea
and cycle on $(1,2,3)$. Note that the evolution equations for
$\Sig_{\alpha}$ and $R^{\alpha}$ decouple from that of
$v^{\alpha}$, discussed below.

It is important to note that if the spatial frame is {\em not\/}
Fermi-propagated $(R^{\alpha} \neq 0)$, the $\Sig_{\alpha\beta}$
evolve in time, with $\Sig^{2} = 1$, both on and off the silent
boundary. On the other hand, if the spatial frame {\em is\/}
Fermi-propagated ($R^{\alpha} = 0$), then $\Sig_{\alpha \beta}$ is
constant in time by Eq.~(\ref{e25}), and diagonal:
\be
\lb{e30}
\Sig_{\alpha \beta} = {\rm diag}\,(\Sig_{1}, \Sig_{2}, \Sig_{3})
\ ,
\ee
with $-2 \leq \Sig_{1} \leq -1 \leq \Sig_{2} \leq 1 \leq \Sig_{3}
\leq 2$ (or permutations thereof). Thus, if the spatial frame is
Fermi-propagated, the Kasner orbits on the silent boundary are
equilibrium points of the shear evolution equations. Since
$\Sig_{\alpha\beta}$ is tracefree and satisfies $\Sig^{2} = 1$, we
obtain
\be
\lb{e31}
\Sig_{1} + \Sig_{2} + \Sig_{3} = 0 \ ,
\ee
and
\be
\lb{e32}
\Sig_{1}^{2} + \Sig_{2}^{2} + \Sig_{3}^{2} = 6 \ .
\ee
The dynamics on the Kasner set also includes the evolution
equations (\ref{e20}) for $v^{\alpha}$ (with $N_{\alpha \beta} =
0$). It follows that if the $\Sig_{\alpha \beta}$ satisfy
Eqs.~(\ref{e30})--(\ref{e32}), then the evolution equations
(\ref{e20}) and (\ref{e21}) for $v^{\alpha}$ admit the equilibrium
sets
$$
v^{\alpha} = 0 \hsp5 {\rm or} \hsp5 v^{2} = 1 \ ,
$$
where the latter condition is also to be supplemented with one of
the six choices for $\vec{v} = (v_{1}, v_{2}, v_{3})^{T}$, namely
\be
\lb{e33}
\vec{v} = \pm\,\vec{E}_{\alpha} \ , \hsp5 \alpha = 1,2,3 \ ,
\ee
where $\vec{E}_{1} = (1,0,0)^{T}$, etc. Thus, there exist {\em
seven\/} sets of equilibrium points forming circles in
$\Sig_{\alpha\beta}$-space, which we shall call {\em Kasner
circles\/} [\,the intersection of the plane~(\ref{e31}) with the
sphere~(\ref{e32})\,], depending on the value of $\vec{v}$ in
Eq.~(\ref{e33}), which we denote by
$$
\cK \ , \hspace{10mm} \cK_{\pm\alpha} \ .
$$
In addition, it follows from Eq.~(\ref{e20}) that for specific
values of $\Sig_{\alpha}$ subject to Eqs.~(\ref{e31}) and
(\ref{e32}), there are six additional {\em lines of equilibrium
points\/}, given by
\be
\lb{e34}
\Sig_{1} =  3\gam-4 \ , \hspace{10mm}  v_{1} > 0
\hsp5 {\rm or} \hsp5 v_{1} < 0 \ , \hspace{10mm}
v_{2} = v_{3} = 0 \ ,
\ee
and cycle on $(1,2,3)$, which join the various Kasner
circles.

At this stage, we digress to describe the symmetry properties of
the Kasner circles. Each circle is divided into six equivalent
sectors,
which we will label according to the ordering of the diagonal shear
components $\Sig_{\alpha}$, which satisfy Eqs.~(\ref{e31}) and
(\ref{e32}). For example, in sector $(123)$ these parameters
satisfy $\Sig_{1} < \Sig_{2} < \Sig_{3}$, etc. The sectors meet at
points where two of the $\Sig_{\alpha}$ are equal. These points are
of two types, conventionally labeled $T_{\alpha}$ (the ``Taub
points'') and $Q_{\alpha}$, given by\footnote{The $T_{\alpha}$
correspond to the Taub form for Minkowski spacetime in the exact
Kasner solution, and the $Q_{\alpha}$ correspond to the locally
rotationally symmetrical non-flat Kasner solution.}
\be
T_{1}: \ (\Sig_{1}, \Sig_{2}, \Sig_{3}) = (2,-1,-1) \ ,
\hspace{10mm}
Q_{1}: \ (\Sig_{1}, \Sig_{2}, \Sig_{3}) = (-2, 1,1) \ ,
\ee
and cycle on $(1,2,3)$. Figure~\ref{fig:kasnc} represents the plane
$\Sig_{1} + \Sig_{2} + \Sig_{3} = 0$ in $\Sig_{\alpha}$-space,
containing a Kasner circle, and showing the six sectors and the
points $T_{\alpha}$ and $Q_{\alpha}$. The figure also shows three
additional points labeled $P_{\alpha}$, which lie outside a
Kasner circle, forming an equilateral triangle whose sides are
tangential to the circle. These points, which are given by
$$
P_{1}: \ (\Sig_{1}, \Sig_{2}, \Sig_{3}) = (-4, 2,2) \ ,
$$
and cycle on $(1,2,3)$, will be used to describe the so-called
curvature transition sets.
\begin{figure}[!htb]
\begin{center}
\includegraphics[scale=0.7]{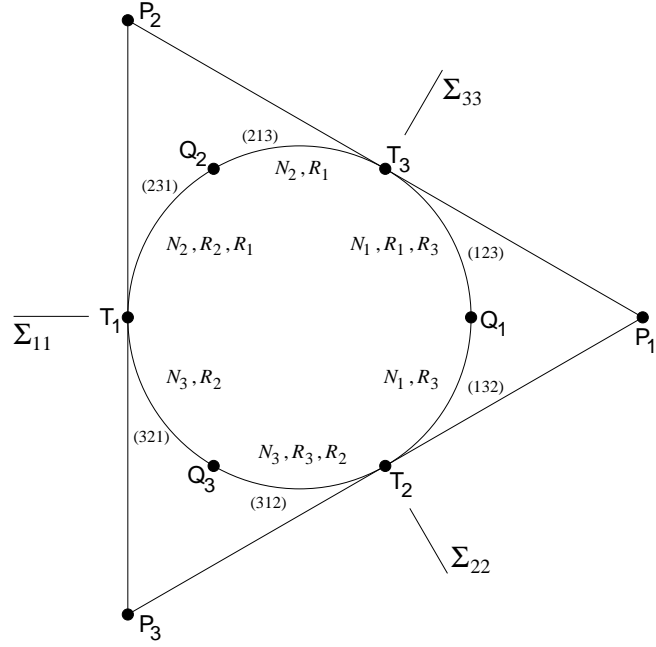}
\end{center}
\caption{A Kasner circle showing the six equivalent sectors and the
variables that are unstable into the past in each sector.}
\lb{fig:kasnc}
\end{figure}
%

\subsection{Transition sets}
The dynamics in the stable subset, defined by Eq.~(\ref{stable}),
is essentially determined by the fact that each Kasner equilibrium
point is a saddle point, with at least two of the nine variables
$N_{\alpha}$, $R^{\alpha}$ and $v^{\alpha}$ being unstable into the
past. Which of these variables are unstable at a particular Kasner
point can be quickly determined by {\em linearizing\/}
Eqs.~(\ref{e14}), (\ref{e15}) and (\ref{e20}) in the neighborhood
of such a point. On the Kasner circle $\cK$ this yields
\bea
\ptl_{t}N_{1} & = & 2(1+\Sig_{1})\,N_{1} \\
\ptl_{t}R_{1} & = & (\Sig_{2}-\Sig_{3})\,R_{1} \\
\ptl_{t}v_{1} & = & (3\gam-4-\Sig_{1})\,v_{1} \ ,
\eea
and similarly for indices~2 and~3. On the Kasner circles $\cK_{\pm
1}$ (where, nearby, $v_{1} = \pm 1 \mp \d v_{1}$, $\d v_{1} > 0$),
the linearized equations for $N_{1}$ and $R_{1}$ remain unchanged,
while Eq.~(\ref{e20}) yields
\bea
\ptl_{t}\d v_{1} & = & -\,2\,\frac{(3\gam-4-\Sig_{1})}{(2-\gam)}
\,\d v_{1} \\
\ptl_{t}v_{2} & = & (\Sig_{1}-\Sig_{2})\,v_{2} \\
\ptl_{t}v_{3} & = & (\Sig_{1}-\Sig_{3})\,v_{3} \ ,
\eea
and similarly for indices~2 and~3 on $\cK_{\pm 2}$ and $\cK_{\pm
  3}$, respectively. It follows that
\bea
N_{1} \quad {\rm is}\;\,  {\rm unstable}\;\, {\rm into}\;\, 
{\rm the}\;\, {\rm past}
&\Leftrightarrow& 1 + \Sig_{1} < 0 \nonumber \\
R_{1} \quad {\rm is}\;\, {\rm unstable}\;\, {\rm into}\;\, 
{\rm the}\;\, {\rm past}
&\Leftrightarrow& \Sig_{2}-\Sig_{3} < 0 \nonumber \\
v_{1} \quad {\rm is}\;\, {\rm unstable}\;\, {\rm into}\;\,
{\rm the}\;\, {\rm past}\;\, {\rm on} \quad \cK
&\Leftrightarrow& 3\gam-4 - \Sig_{1} < 0 \nonumber \\
v_{1} \quad {\rm is}\;\, {\rm unstable}\;\, {\rm into}\;\, 
{\rm the}\;\, {\rm past}\;\, {\rm on} \quad
\cK_{\pm 1} &\Leftrightarrow& 3\gam-4 - \Sig_{1} > 0 \nonumber \\
v_{2} \quad {\rm is}\;\, {\rm unstable}\;\, {\rm into}\;\,
{\rm the}\;\, {\rm past}\;\, {\rm on} \quad \cK_{\pm 1}
&\Leftrightarrow& \Sig_{1}-\Sig_{2} < 0 \nonumber \\
v_{3} \quad {\rm is}\;\, {\rm unstable}\;\, {\rm into}\;\,
{\rm the}\;\, {\rm past}\;\, {\rm on} \quad \cK_{\pm 1}
&\Leftrightarrow& \Sig_{1}-\Sig_{3} < 0 \nonumber \ ,
\eea
and similarly for indices~2 and~3. The arcs of the Kasner circles
on which the variables $N_{\alpha}$ and $R^{\alpha}$ are unstable
are shown in Fig.~\ref{fig:kasnc}, and those on which the variables
$v^{\alpha}$ are unstable are shown in Figs.~\ref{fig:kasncv}
and~\ref{fig:kasncexv}.
\begin{figure}[!htb]
\begin{center}
\includegraphics[scale=0.7]{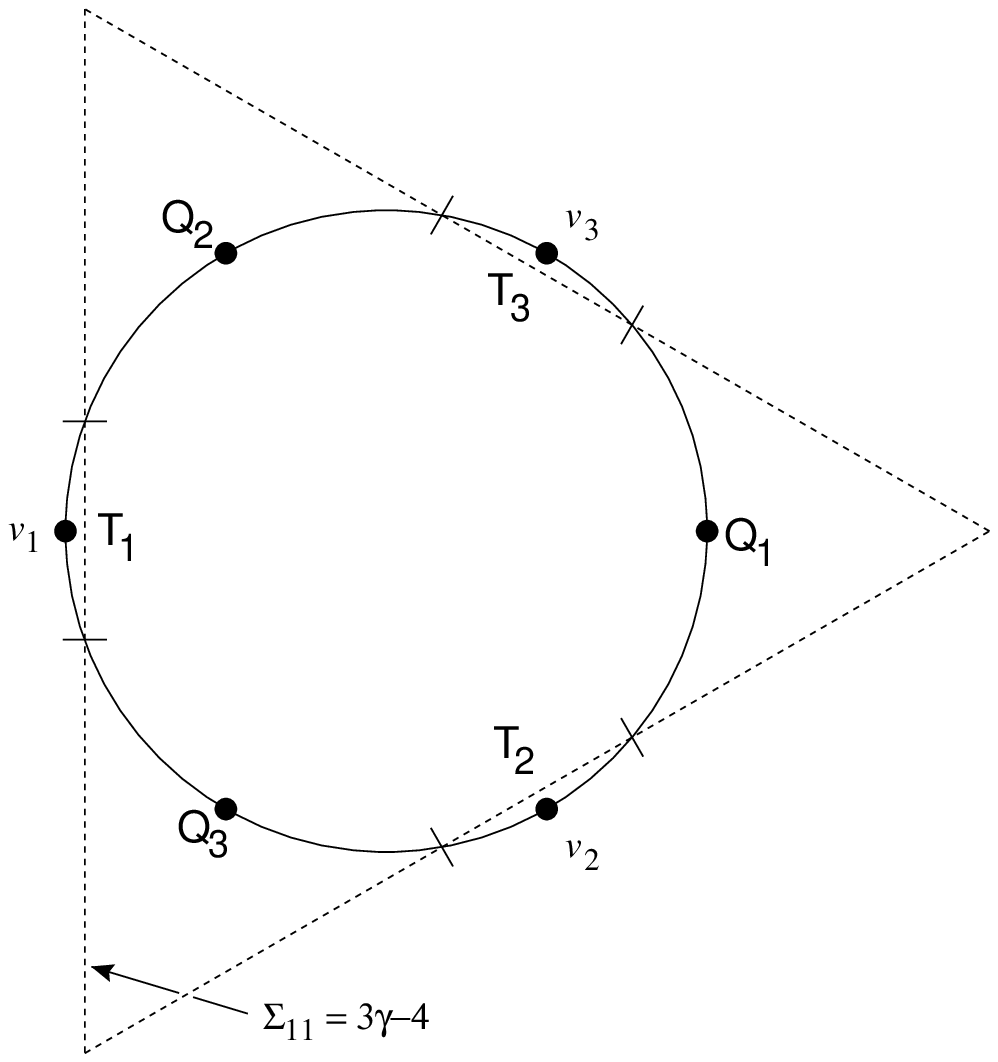}  \hspace*{1in}
\includegraphics[scale=0.7]{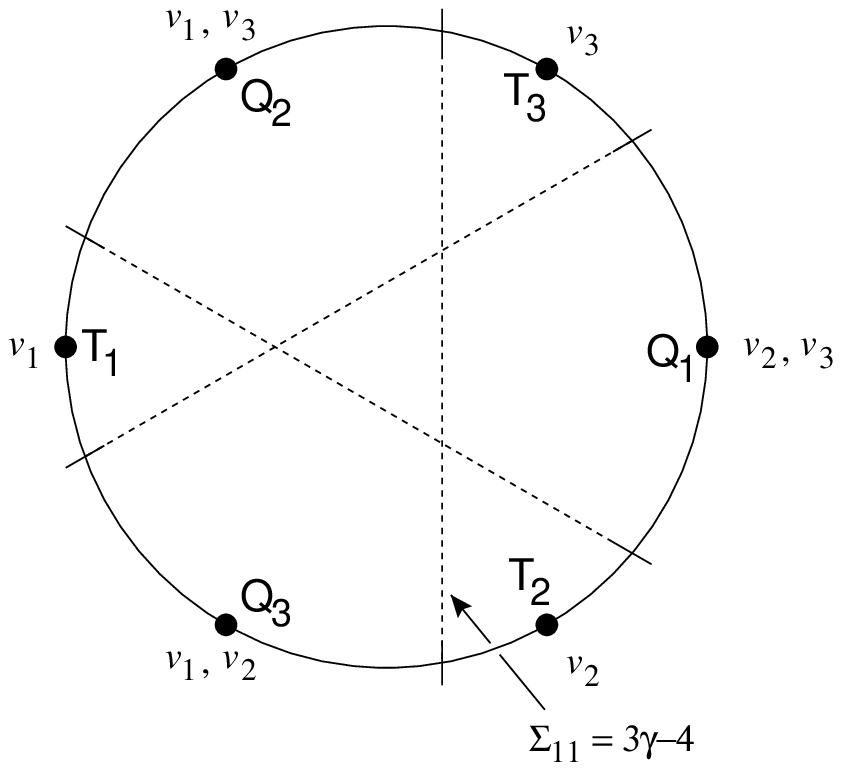}
\end{center}
\caption{The Kasner circle $\cK$ showing the arcs on which the
variables $v^{\alpha}$ are unstable into the past. Figure 2a
shows the case $\sfrac{5}{3} < \gam < 2$ and figure 2b the case $1
< \gam < \sfrac{5}{3}$.}
\lb{fig:kasncv}
\end{figure}
\begin{figure}[!htb]
\begin{center}
\includegraphics[scale=0.7]{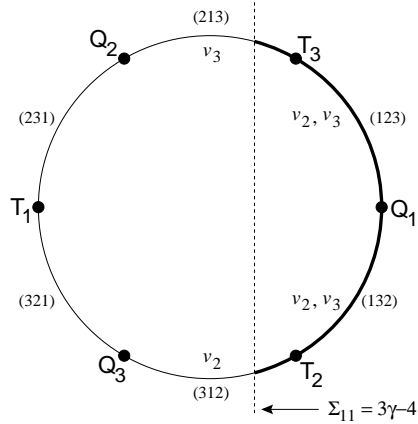}
\end{center}
\caption{The Kasner circles $\cK_{\pm 1}$ showing the arcs on
which the variables $v^{\alpha}$ are unstable into the past. The
variable $v_{1}$ is unstable on the boldface arc to the right of
the line $\Sig_{11} = 3\gam-4$.}
\lb{fig:kasncexv}
\end{figure}
A given Kasner equilibrium point $P$ thus has an unstable manifold
(into the past).\footnote{As Figs.~\ref{fig:kasnc}--\ref{fig:kasncexv}
show, the unstable manifold is at most six-dimensional.} The key
property of this unstable manifold is that {\em each of its orbits
join $P$ to some other Kasner point\/}. The simplest such orbits
are those on which only one of the nine variables $N_{\alpha}$,
$R^{\alpha}$ and $v^{\alpha}$ is non-zero, or those on which two
$v^{\alpha}$ are non-zero but with extreme tilt $(v^{2} =
1)$. These special orbits, which we shall refer to collectively as
{\em transition sets\/}, are listed by name and symbol in
Tab.~1. In this table the subscripted letter on the $\cT$ indicates
the ``excited'' variable.
\begin{table}[!htb]
\caption{The transition sets}

\begin{center}
\begin{tabular}{p{1.25in}p{1.2in}l}
{\em Name} & {\em Symbol} 
&  {\em Transitions}\\
\hline &  &  $\cK \rightarrow \cK \quad 
(v^{\alpha} = 0)$\\
curvature & $\cT_{N_{\alpha}}$ & or \\
& &   $\cK_{\pm \alpha} \rightarrow \cK_{\pm \alpha}
\quad (v^{2} =1)$ \\[1mm] \hline
&  &  $\cK \rightarrow \cK \quad
(v^{\alpha} = 0)$\\
frame & $\cT_{R_{\alpha}}$ & or \\
&&   $\cK_{\pm \alpha} \rightarrow \cK_{\pm \alpha} \quad
(v^{2} = 1)$ \\[1mm] \hline
tilt & $\cT_{v_{\alpha}}$ & $\cK \rightarrow 
\cK_{\pm \alpha}$ \quad or vice versa \\
& &  $(\Sigma_{\alpha \beta}$ fixed)\\[1mm] \hline
extreme-tilt & $\cT_{v_{\alpha} v_{\beta}}$ &  
$\cK_{\pm \alpha} \rightarrow \cK_{\pm \beta}$ 
\\ 
& $(\alpha \neq \beta)$ & $(\Sigma_{\alpha \beta}$
fixed, $v^{2} =1$) \\[1mm] \hline
\end{tabular}
\end{center}
\lb{tab:trset}
\end{table}
We now discuss these transition sets in turn. \enl

Firstly, there are the {\em curvature transition sets\/}
$\cT_{N_{\alpha}}$, $\alpha = 1,2,3,$ which are shown in
Fig.~\ref{fig:ctrans}. In this and all subsequent figures, {\em
orbits are directed toward the past\/}. For $\alpha = 1$ the
curvature transition orbits are given by
\be
\lb{curvtr}
(1-k)\,(2-\Sig_{2}) = (1+k)\,(2-\Sig_{3}) \ ,
\ee
where $k$ is a parameter that satisfies $-1 \leq k \leq 1$. This
relation follows from Eqs.~(\ref{e16}) and~(\ref{e14}) with $0 = N_{2}
= N_{3} = R^{\alpha}$. In the spatially homogeneous setting these
orbits describe the Taub vacuum Bianchi Type--II solutions, and
determine the past attractor for vacuum and non-tilted SH models of
Bianchi Type--VIII and Type--IX (in a Fermi-propagated frame; see
WE, Fig.~6.6 on p.~138, and p.~143--7, but note differences in
labeling). These curvature transitions link different ``Kasner
epochs'', according to a transition law for the Kasner exponents
first found by BKL (see Ref.~\ct{bkl70}, p.~535--7; also WE,
p.~236). We derive this transition law in
App.~\ref{subsec:curvtr}. \enl
\begin{figure}[!htb]
\begin{center}
\includegraphics[scale=0.5]{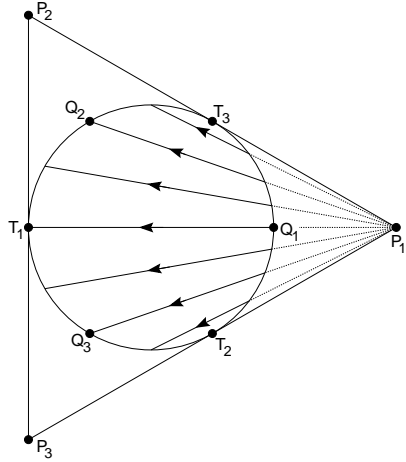}
\end{center}
\caption{The curvature transition set $\cT_{N_{1}}$.  The sets
$\cT_{N_{2}}$ and $\cT_{N_{3}}$ are obtained by cycling on
$(1,2,3)$.}
\lb{fig:ctrans}
\end{figure}

Secondly, there are the {\em frame transition sets},
$\cT_{R_{\alpha}}$, which are shown in Fig.~\ref{fig:ftrans}. For
$\alpha = 1$ they are given by
$$
\Sig_{1} = k \ ,
$$
where $k$ is a parameter that satisfies $-2 < k < 2$. In the
spatially homogeneous setting these transition sets map a Kasner
solution into a physically equivalent Kasner solution through
rotation of the spatial frame by $\pi/2$ about one of its axes.
For example, the transition sets $\cT_{R_{1}}$ result in the
interchange $\Sig_{2} < \Sig_{3} \rightarrow \Sig_{3} < \Sig_{2}$,
as can be seen by comparing Figs.~\ref{fig:kasnc}
and~\ref{fig:ftrans}a. \enl
\begin{figure}[!htb]
\begin{center}
\includegraphics[scale=0.5]{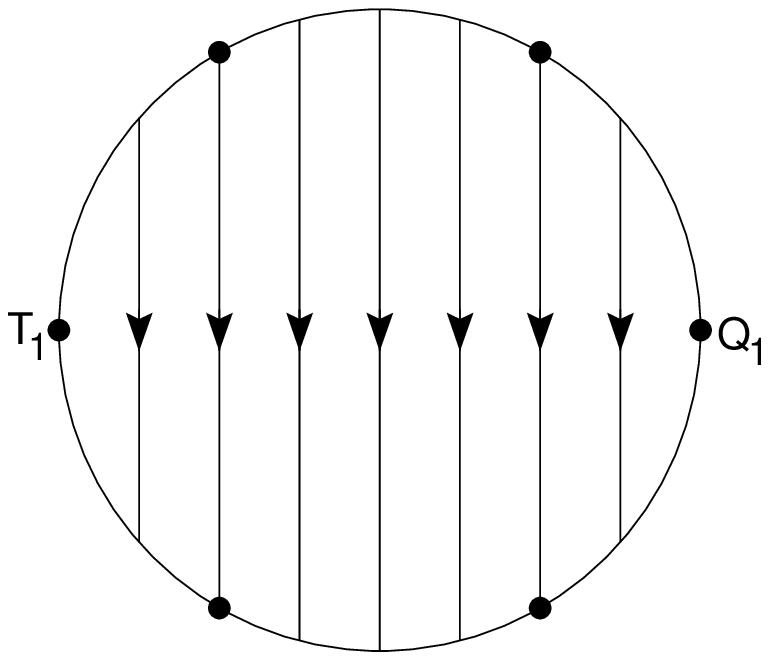} \qquad
\includegraphics[scale=0.5]{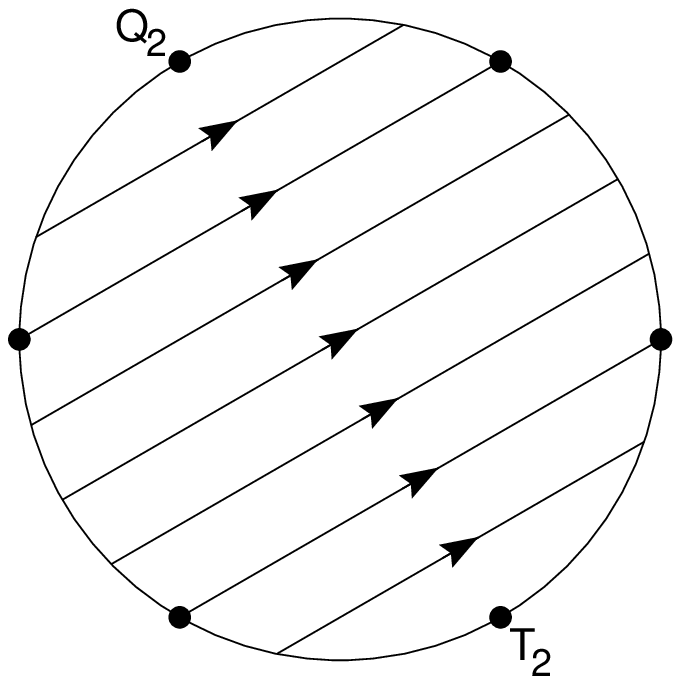} \qquad
\includegraphics[scale=0.5]{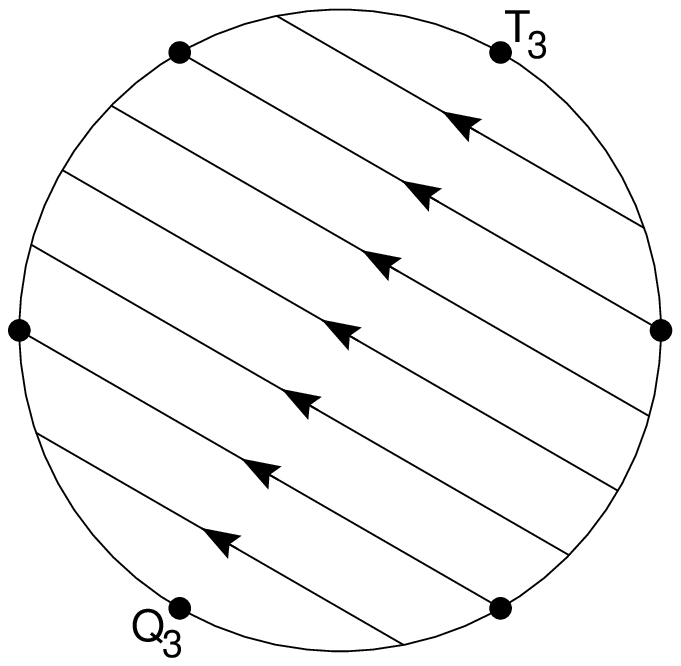}
\end{center}
\caption{The frame transitions sets $\cT_{R_{1}}$  (5a),
  $\cT_{R_{2}}$ (5b) and $\cT_{R_{3}}$ (5c).}
\lb{fig:ftrans}
\end{figure}

Thirdly, there are the {\em tilt transition sets\/},
$\cT_{v_{\alpha}}$, which are shown in Fig.~\ref{fig:ttrans}. They
are simply lines with $\Sig_{\alpha}$ constant and one of the
$v^{\alpha}$ non-zero. Whether the orbits join a point on $\cK$ to
a point on one of the Kasner circles $\cK_{\pm \alpha}$, or vice
versa, depends on the values of the $\Sigma_{\alpha}$ (see
Figs.~\ref{fig:kasncv} and \ref{fig:kasncexv}). The reversal of
direction of these orbits is governed by the six lines of
equilibrium points given by Eq.~(\ref{e34}). \enl
\begin{figure}[!htb]
\begin{center}
\includegraphics[scale=0.7]{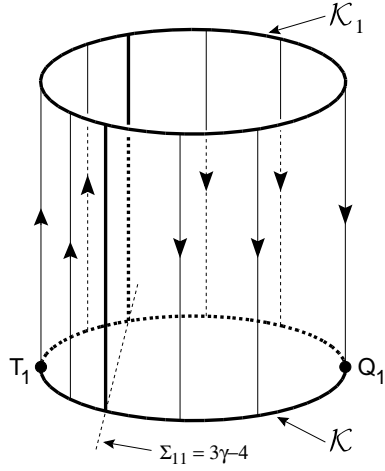}
\end{center}
\caption{The tilt transition set $\cT_{v_{1}}$.}
\lb{fig:ttrans}
\end{figure}

Finally, there are the {\em extreme-tilt transition sets\/}. Let
us consider the subset $v^{2} = 1$ (with fixed $\Sig_{\alpha}$).
Then $v^{\alpha} = e^{\alpha}$, where $e^{\alpha}$ is a unit
vector, which may be parametrized according to
\be
e^{\alpha} = (\cos\vartheta, \sin\vartheta\cos\varphi,
\sin\vartheta\sin\varphi)^{T} \ ,
\hspace{5mm}
0 \leq \vartheta \leq \pi \ ,
\hspace{5mm}
0 \leq \varphi \leq 2\pi \ .
\ee
We then obtain from Eq.~(\ref{e20}) a simple dynamical system
for the polar angles $\{\vartheta, \varphi\}$, given by
\bea
\begin{array}{lclclcl}
\ptl_{t}\vartheta & = & -\,a^{2}\,\sin 2\vartheta & \hsp5 &
a^{2} & := & \sfrac{3}{2}\,(-\,p_{1}+p_{2}\cos^{2}\varphi
+p_{3}\sin^{2}\varphi) \\
\ptl_{t}\varphi & = & -\,b^{2}\,\sin 2\varphi & \hsp5 &
b^{2} & := & -\,\sfrac{3}{2}\,(p_{2}-p_{3})
\end{array} \ .
\eea
It is easily seen that if $\Sig_{1} < \Sig_{2} < \Sig_{3}$ (or
equivalently $p_1<p_2<p_3$), then the past attractor of this
dynamical system is given by $\{\vartheta, \varphi\} =
\{\sfrac{1}{2}\pi, \sfrac{1}{2}\pi\}$ and $\{\vartheta, \varphi\} =
\{\sfrac{1}{2}\pi, \sfrac{3}{2}\pi\}$, and, hence,
\be
\lim_{t\rightarrow -\infty} e^{\alpha} = \pm\,\vec{E}_{3} \ ;
\ee
similarly for other orderings of $\Sig_{\alpha}$. Thus,
extreme-tilt transition sets are orbits that lie on the
extreme-tilt sphere $v^{2} = 1$ in $v^{\alpha}$-space, with $0 =
N_{\alpha} = R^{\alpha}$, and $\Sig_{\alpha}$ fixed.
Figure~\ref{fig:ettrans} shows the directions corresponding to
$\Sig_{1} < \Sig_{2} < \Sig_{3}$, i.e., the arc $(123)$ on the
Kasner circles. The other cases can be obtained by interchanging 1,
2 and 3. The points $B_{\pm\alpha}$ in Fig.~\ref{fig:ettrans}
correspond to the points on the Kasner circles $\cK_{\pm\alpha}$
determined by the values of the $\Sig_{\alpha}$. The directions of
the orbits joining the points $B_{\pm\alpha}$ depend on the
ordering of the $\Sig_{\alpha}$.
\begin{figure}[!htb]
\begin{center}
\includegraphics[scale=0.75]{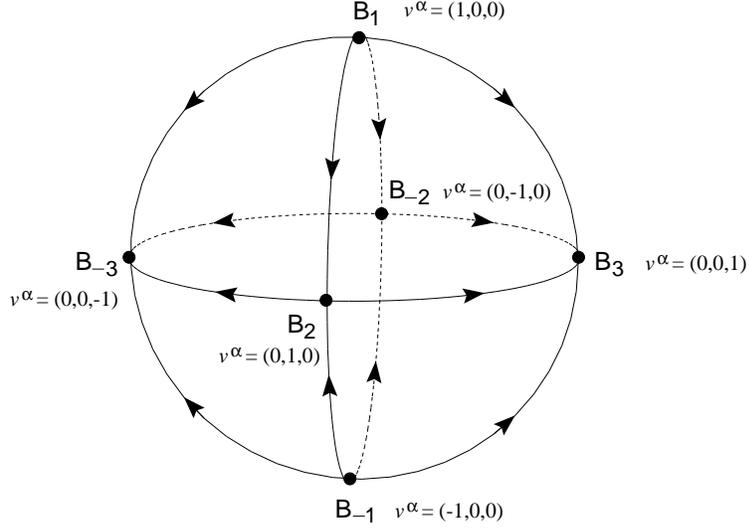}
\end{center}
\caption{The extreme-tilt transition sets on the extreme-tilt
sphere $v^{2} = 1$, $0 = N_{\alpha} = R^{\alpha}$, with
$\Sig_{\alpha}$ fixed and $\Sig_{1} < \Sig_{2} < \Sig_{3}$.}
\lb{fig:ettrans}
\end{figure}
%

\subsection{Structure of the past attractor}
\lb{subsec:pastruc}
We are now in a position to state our conjecture concerning the
local past attractor of the $G_{0}$ evolution equations for generic
initial data satisfying the constraints. We introduce the following
notation:
\be
\cT_{N}= \cup\,\cT_{N_{\alpha}} \ , \hsp5
\cT_{R} = \cup\,\cT_{R_{\alpha}} \ , \hsp5
\cT_{\rm tilt} = \cup\,\cT_{v_{\alpha}} \ , \hsp5
\cT_{\rm extreme} = \cup\,\cT_{v_{\alpha}v_{\beta}} \ , \hsp5
\cK_{\rm extreme} = \cup\,\cK_{\pm\alpha} \ .
\ee
Thus, $\cT_{N}$ is the union of all curvature transition sets,
$\cT_{R}$ is the union of all frame transition sets, etc. (see
Tab.~1 for the complete list of transition sets).

We now make the following conjecture concerning the local past
attractor $\cA^{-}$. \enl

\hangindent=9em
{\bf Conjecture 3}:
The local past attractor $\cA^{-}$ for $G_{0}$ cosmologies with a
silent initial singularity is given by
\be
\lb{e35}
\cA^{-} = \cK \cup \cK_{\rm extreme} \cup \cT_{N} \cup
\cT_{R} \cup \cT_{\rm tilt} \cup \cT_{\rm extreme} \ .
\ee
\enl

\noindent
The essential property of the various transition sets is that they
define so-called {\em infinite heteroclinic sequences\/} on the
past attractor, i.e., infinite sequences of Kasner equilibrium
points joined by transition sets, oriented into the past.  In
particular, a typical Kasner point will be the starting point for
infinitely many heteroclinic sequences (infinitely many, because at
least two transition sets (three, when $\gam < \sfrac{5}{3}$)
emanate from each Kasner point). The significance of the
heteroclinic sequences is that an orbit that is asymptotic to the
past attractor (for given values of the $x^{i}$) will shadow a
heteroclinic sequence, and hence the cosmological model will be
approximated locally by a sequence of Kasner states.

The conjectured past attractor, $\cA^{-}$, is a proper subset of
the stable subset, defined by Eq.~(\ref{stable}). Thus, in addition
to the stable variables (\ref{e8}), various expressions involving
the unstable variables $N_{\alpha}$, $R^{\alpha}$ and $v^{\alpha}$
will tend to zero on the attractor, even though the limits of these
variables as $t \rightarrow -\infty$ do not exist. The desired
expressions depend on the following quantities:
$$
N^{2} := N_{\alpha}N^{\alpha} \ , \hspace{10mm}
R^{2} := R_{\alpha}R^{\alpha} \ , \hspace{10mm}
v^{2} := v_{\alpha} v^{\alpha} \ ,
$$
and
$$
\Delta_{N} := (N_{1}N_{2})^{2} + (N_{2}N_{3})^{2} +
(N_{3}N_{1})^{2} \ , 
$$
together with analogous quantities $\Delta_{R}$ and $\Delta_{v}$
for $R^{\alpha}$ and $v^{\alpha}$. The past attractor is then
characterized by the following limits [\,in addition to
Eqs.~(\ref{e82})--(\ref{e83}) and (\ref{e10})\,]:
\bea
\lb{e37}
& \lim_{t \rightarrow -\infty} & N^{2}R^{2} = 0 \\
\lb{e38}
& \lim_{t \rightarrow -\infty} & v^{2}\,(1- v^{2})\,
(N^{2}+R^{2}) = 0 \\
\lb{e39}
& \lim_{t \rightarrow -\infty} & (\Delta_{N}, \Delta_{R})
= (0,0) \\
\lb{e40}
& \lim_{t \rightarrow -\infty} & \left(\Delta_{v}\,
(N^{2}+R^{2}), \,\Delta_{v}\,(1-v^{2})\right) = (0,0) \\
\lb{e41}
& \lim_{t \rightarrow -\infty} & v_{1}v_{2}v_{3} = 0 \ .
\eea
The limits~(\ref{e37}) and (\ref{e38}) imply that if $N_{\alpha}$
($R^{\alpha}$, respectively) is active, than $R^{\alpha}$
($N_{\alpha}$, respectively) must be close to zero, and likewise
either $v^{2}$ or $1-v^{2}$ must be close to zero, as on the
$\cT_{N_{\alpha}} (\cT_{R_{\alpha}})$ transition sets,
respectively. The limit (\ref{e38}) also implies that if $v^{2}$ is
not close to 0 or 1, then $N_{\alpha}$ and $R^{\alpha}$ must be
close to zero, as on the tilt transition sets $\cT_{v_{\alpha}}$.
The limit~(\ref{e39}) implies that at most one $N_{\alpha}$ and at
most one $R^{\alpha}$ can be active simultaneously, as on the
$\cT_{N_{\alpha}}$ and $\cT_{R_{\alpha}}$ transition sets. The
limit~(\ref{e40}) implies that if two $v^{\alpha}$ are active
simultaneously, (i.e., $\Delta_{v} \neq 0$), then $N_{\alpha}$ and
$R^{\alpha}$ must be close to zero and $v^{2}$ must be close to 1,
as on the extreme-tilt transition sets $\cT_{v_{\alpha}v_{\beta}}$.

The conjectured structure of the past attractor $\cA^{-}$ in terms
of Kasner equilibrium points and transition sets, as given by
Eq.~(\ref{e35}), or as described by the
limits~(\ref{e82})--(\ref{e83}), (\ref{e10}) and
(\ref{e37})--(\ref{e41}), embodies the notion that as one
approaches the attractor into the past along an orbit, the
probability that more than one of the nine unstable variables
$N_{\alpha}$, $R^{\alpha}$ and $v^{\alpha}$ is active during any
one transition (except for the extreme-tilt transition sets on
which two $v^{\alpha}$ are non-zero, but are constrained by $v^{2}
= 1$) tends to zero. Evidence that the probability of multiple
transitions involving pairs such as $(R_{1},N_{1})$ or
$(R_{1},R_{2})$ tends to zero as $t \rightarrow - \infty$ is
provided by numerical simulations for non-tilted SH models of
Bianchi Type--VI$^{*}_{-1/9}$, reported by Hewitt {\em et
al\/}~\ct{hewetal2002}.

We shall refer to the local past attractor $\mathcal{A}^{-}$,
defined by Eq.~(\ref{e35}), as the {\em generalized Mixmaster
attractor\/}, since it generalizes the past attractor for the
so-called Mixmaster models (SH models of Bianchi Type--IX; see WE,
p.~146, and Tab.~2 to follow), making precise the heuristic
notion of an oscillatory singularity.

\section{Cosmologies with isometries}
\lb{sec:cossym}
In Sec.~\ref{sec:att}, we proposed a detailed structure for the past
attractor for $G_{0}$ cosmologies with a silent initial singularity
[\,see Eq.~(\ref{e35})\,]. Classes of cosmologies that admit an
isometry of some sort are described by invariant sets of the
Hubble-normalized state space.  In this section we exploit this
fact to predict the structure of the past attractor for these more
specialized models, thereby providing a link to much recent
research.

For models with an isometry it is possible that one or more of the
nine unstable variables $N_{\alpha}$, $R^{\alpha}$ or $v^{\alpha}$
is required to be zero, leading to two possibilities.

\begin{itemize}

\item[(i)] The initial singularity is {\em oscillatory\/}.

This possibility occurs if each arc of the various Kasner circles
has at least one unstable variable (refer to
Figs.~\ref{fig:kasnc}--\ref{fig:kasncexv}). The attractor will then
include all available transition sets, and the evolution into the
past along a typical timeline will be described by an infinite
sequence of Kasner states, possibly of a more specialized nature
than for $G_{0}$ cosmologies.

\item[(ii)] The initial singularity is {\em Kasner-like\/}.

This possibility occurs if at least one arc on one of the Kasner
circles has no unstable variables. The arc(s) in question then form
the past attractor, and the evolution into the past along a typical
timeline will be described by a specific Kasner state. A
cosmological solution with this type of singularity is also
referred to as being {\em asymptotically
velocity(-term)-dominated\/}, a term that has its origins in the
work of Eardley, Liang and Sachs~\ct{earetal72} and Isenberg and
Moncrief~\ct{isen90}.

\end{itemize}

We now present various classes of cosmologies with isometries,
whose initial singularities have been discussed in the literature,
and give the conjectured past attractor in terms of our
formulation. The specific nature of the isometry determines whether
the initial singularity is oscillatory or Kasner-like.

Firstly, we consider SH cosmologies, which, as we have seen in
Subsec.~\ref{subsec:shcosm}, can be described by the
finite-dimensional reduced state space defined by the variables
$\vec{Y}$.  Since the definition of the generalized Mixmaster
attractor $\cA^{-}$ involves only the variables $\vec{Y}$, the
attractor also exists as an invariant set in this reduced state
space.  Indeed, we conjecture that in this context $\cA^{-}$ {\em
is the past attractor for the general class of SH cosmologies}, and
that it will thus contain the past attractors for the three special
classes of SH cosmologies with an oscillatory singularity that have
been analyzed in detail to date. In Tab.~2 we give these three
special classes of SH cosmologies and list the key
references. These papers use Hubble-normalized variables, but there
are some differences in the labeling of variables compared to the
present paper. It is notoriously difficult to prove rigorous
results about oscillatory singularities and little progress has
been made until recently, when Ringstr\"{o}m~\ct{rin2001}, in a
remarkable piece of mathematical analysis, rigorously established
the existence of the past attractor for the class of non-tilted
Bianchi Type--IX cosmologies, by proving the required limits,
conjectured earlier by WE (see p.~146--7).

\begin{table}[h]
\caption{Perfect fluid SH cosmologies with oscillatory initial
singularity}\

\medskip
\begin{center}
\begin{tabular}{ccc} \hline \\
{\em Class of models\/} & {\em Non-zero\/} & {\em Past attractor\/}\\
& {\em unstable variables\/} & \\[2mm] \hline 
&&\\
non-tilted Type--VIII and Type--IX & $N_{1}, N_{2}, N_{3}$ & $\cK \cup
\cT_{N}$\\
(WE, p.~146, Ringstr\"{o}m~\ct{rin2001}) & & \\[4mm]
non-tilted Type--VI$^{*}_{-1/9}$ & $N_{3}, R_{1}, R_{3}$ & $\cK \cup
\cT_{N_{3}} \cup \cT_{R_{1}} \cup \cT_{R_{3}}$\\
(Hewitt {\em et al\/}~\ct{hewetal2002}) & & \\[4mm]
tilted Type--II & $N_{3}, R_{1}, R_{3}, v_{1}$ & $\cK \cup \cK_{1} \cup
\cT_{N_{3}} \cup \cT_{R_{1}} \cup \cT_{R_{3}} \cup \cT_{v_{1}}$\\
(Hewitt {\em et al\/}~\ct{hewetal2001}) & &\\ \hline 
\end{tabular}
\end{center}
\end{table}

Secondly we consider spatially inhomogeneous cosmologies. Most
recent research on the initial singularity has been restricted in
two ways:

\begin{itemize}

\item[(i)]
the spacetime is assumed to have  compact spatial sections,

\item[(ii)]
the energy--momentum--stress tensor is assumed to be zero (vacuum
solutions).
\end{itemize}

\noindent
The first restriction is made because it enables one to prove
results about the global existence of solutions. It does not,
however, affect the structure of the past attractor, since it is
determined by the dynamics along individual timelines. In view of
the BKL conjecture, namely that matter is not significant
dynamically as the initial singularity is approached, one might
believe that the second restriction can be made without loss of
generality when determining the past attractor. This conclusion is
not valid, however. Our analysis leads to the conjecture that the
past attractor for vacuum $G_{0}$ models is in fact the much
simpler set given by
\be
\lb{e141}
\cA_{\rm vac}^{-} = \cK \cup \cT_{N} \cup \cT_{R} \ ,
\ee
since the Hubble-normalized state vector for vacuum models does not
contain the peculiar velocity variable $v^{\alpha}$, which implies
that the extreme Kasner circles $\cK_{\rm extreme}$ and the
transition sets $\cT_{\rm tilt}$ and $\cT_{\rm extreme}$ cannot be
part of the past attractor [\,see Eq.~(\ref{e35})\,]. Nevertheless,
determining the vacuum past attractor is an important first step in
determining the attractor for non-vacuum models.

In Tab.~3 we list the classes of vacuum spatially inhomogeneous
cosmologies whose initial singularity has been studied.  In each
case we can predict immediately whether the singularity will be
oscillatory or Kasner-like. In the table we give the past attractor
for each class, which is a subset of the general vacuum attractor
$\cA_{\rm vac}^{-}$.\footnote{At this stage the reader might be
concerned with the fact that these models have not been studied in
the separable volume gauge. However, we believe that, due to
asymptotic silence, our discussion is ``gauge robust'', i.e, that
the local asymptotic temporal behavior is not affected by the
choice of temporal gauge. To make this more substantial we note
that the choice $\cn = 1$ and $N^{i} = 0$ was not necessary for
obtaining our picture of the past attractor. Any sufficiently
smooth choice $\cn(\vec{X})$ such that $\cn$ is positive and
bounded on the attractor does not change the flow on the past
attractor and thus one would obtain the same results as the choice
$\cn = 1$ yields; $N^{i}$ can be similarly generalized. We also
note that these are not necessary conditions, and that even wider
sets of gauge choices are allowed if one takes into account the
detailed structure of the EFE.}

\begin{table}[h]
\caption{Past attractor for vacuum spatially inhomogeneous
cosmologies with isometries. In the first three cases, the initial
singularity is Kasner-like and in the last two cases, the initial
singularity is oscillatory.}

\smallskip
\begin{center}

\begin{tabular}{ccc}\hline
&&\\
{\em Class of models\/} & {\em Non-zero\/} & {\em Past vacuum\/} \\
& {\em unstable variables\/} & {\em attractor\/}\\ \hline
&&\\
Polarized Gowdy & all zero & $\cK$\\
$\equiv$ diagonal $G_{2}$ &&\\
(Isenberg and Moncrief~\ct{isen90} ) &&\\[3mm]
Unpolarized Gowdy & $N_{3}, R_{1}$ & arc$(T_{2}Q_{1}) \subset \cK$\\
$\equiv$ OT $G_{2}$ &&\\
(Kichenassamy and Rendall~\ct{kicren98})\\[3mm]
Polarized $T_{2}$-symmetric & $R_{3}$ & arc$(T_{3} Q_{2} T_{1} Q_{3})
\subset \cK$\\
$\equiv G_{2}$ with one HO KVF &&\\
(Isenberg and Kichenassamy~\ct{isen99})\\[3mm]
$T_{2}$-symmetric & $N_{3}, R_{1}, R_{3}$ & $\cK \cup \cT_{N_{3}} \cup
\cT_{R_{1}} \cup \cT_{R_{3}}$ \\
$\equiv$ generic $G_{2}$ && \\
(Berger {\em et al\/}~\ct{berg01}) \\[3mm]
$U(1)$-symmetric & all non-zero & $\cK \cup \cT_{N} \cup \cT_{R}$\\
$\equiv$ generic $G_{1}$ &&\\
(Berger and Moncrief~\ct{bergmon98b}) && \\[3mm] \hline
\end{tabular}
\end{center}
\end{table}

One class of vacuum models is not included in Tab.~3, the so-called
polarized $U(1)$-symmetric models (Berger and
Moncrief~\ct{bergmon98a}). These are $G_{1}$ cosmologies for which
the single spacelike Killing vector field is
hypersurface-orthogonal. The reason for this exclusion is that the
dynamical consequences of the hypersurface-orthogonality condition
are not compatible with our choice of spatial frame, given by
Eq.~(\ref{frame}). These models could be incorporated by making a
different choice of spatial frame, as discussed in footnote~10, but
we will not pursue this matter here.

It should be noted that in the papers listed in Tab.~3 the
conclusions about the dynamics near the initial singularity are not
expressed in terms of a past attractor: we have reformulated their
results within our dynamical systems framework, and at this stage
most of the results about the past attractor have not been
rigorously established. The papers referred to use a metric-based
approach\footnote{In a recent paper~\ct{andetal2002}, however, the
Gowdy models are analyzed using scale-invariant variables
introduced in Refs.~\ct{hewwai90} and~\ct{hveetal2002}.} instead of
the orthonormal frame approach. Some of them make use of the
so-called Fuchsian algorithm to establish the asymptotics at a
Kasner-like initial singularity (see Refs.~\ct{kicren98}
and~\ct{isen99}) while others rely on a Hamiltonian formalism and
the so-called method of consistent potentials to predict whether
the initial singularity will be oscillatory or not (see
Refs.~\ct{bergmon98b}, \ct{berg98} and~\ct{berg01}). In this
approach, the transitions between Kasner states are described
heuristically as bounces off potential walls determined by the
Hamiltonian. Some of these papers also describe numerical
simulations that display a finite number of Mixmaster oscillations.

\section{Concluding remarks}
\lb{sec:disc}
In this paper we have developed a mathematical framework for
analyzing the dynamics of $G_{0}$ cosmologies, and in particular
the BKL conjecture discussed in the Introduction. A key step was
the introduction of scale-invariant variables, using the Hubble
scalar defined by a timelike reference congruence as the
normalization factor. One of the principal advantages of
Hubble-normalization lies in the behavior of the dynamical
variables as the initial singularity is approached: the dimensional
variables diverge, while, for at least a generic family of
solutions, {\em the Hubble-normalized variables remain bounded\/}.

The structure of the evolution equations and constraints led to the
introduction of the {\em silent boundary\/} in the
Hubble-normalized state space and enabled us to define a {\em
silent initial singularity\/}. The next step was to construct the
{\em generalized Mixmaster attractor\/}, which makes precise the
notion of an oscillatory initial singularity in a $G_{0}$
cosmology, while having a simple geometrical structure (see
Figs.~\ref{fig:kasnc}--\ref{fig:ettrans}), and allowed three precise
conjectures on early cosmological dynamics to be formulated
(Conjecture~1 in Subsec.~\ref{subsec:silinising}, Conjecture~2 in
Subsec.~\ref{subsec:stabset}, and Conjecture~3 in
Subsec.~\ref{subsec:pastruc}). The construction of the past attractor
also highlights and clarifies the important r\^{o}le of SH dynamics
in the $G_{0}$ context. Indeed, there is now considerable evidence,
both numerical and analytical, that SH dynamics influences the
asymptotic dynamics of spatially inhomogeneous cosmologies near the
initial singularity in a significant way. Our formulation places
this relationship on a sound footing: \textit{the local past
attractor for\/} $G_{0}$ \textit{cosmologies with a silent initial
singularity is the past attractor for SH cosmologies\/}. We are now
in a position to restate the BKL conjecture in a precise form:

\pagebreak
\begin{quote}
For almost all cosmological solutions of Einstein's field
equations, a spacelike initial singularity is {\em silent,
vacuum-dominated\/} and {\em oscillatory\/}.
\end{quote}

\noindent
Proving this conjecture entails establishing all the limits
associated with Conjectures~1, 2, and 3 in Sec.~\ref{sec:att}. As a
first step, one would have to complete the proof for the SH models,
begun by Ringstr\"{o}m~\ct{rin2001}. A natural second step would be
to consider the simplest class of spatially inhomogeneous models
with an oscillatory initial singularity, namely the generic $G_2$
models (see Tab.~3), restricting consideration to vacuum solutions
for simplicity. Analyzing the r\^{o}le of the spatial derivatives
in a neighborhood of the silent boundary will be a major step in
this analysis, and will clearly present a formidable challenge.

This unifying statement incorporates certain fundamental physical
ideas about singularities, partially supported by known examples
and theorems. It is useful to revisit the conjectured physical
behavior in a way that highlights various aspects of the
situation:

\begin{itemize}
\item[(i)] The generic cosmological initial singularity is a
strong-gravity phenomenon, and so should be linked to trapped
surfaces, which intuitively capture the notion of ultra-strong
gravitational fields (and thus also to the standard singularity
theorems).

\item[(ii)] The generic cosmological initial singularity is likely
to be a spacelike curvature singularity because a null singularity
will be very special and timelike singularities will by their
nature intersect relatively few worldlines of matter (but
confirming this will depend on implementing a good measure on the
space of cosmological models, which is needed in any case in order
to put on a firm footing all talk about probabilities).

\item[(iii)] The generic spacelike curvature singularity is a
scalar curvature singularity, since non-scalar curvature
singularities require fine tuning of initial data.

\item[(iv)] If the energy conditions are strictly obeyed, the
curvature singularity is generically Weyl curvature dominated, at
least when vorticity in the matter fluid is not significant (this
is not the case if the energy conditions are just marginally
satisfied, as exemplified by stiff perfect fluids, but these are
not physically likely states).

\item[(v)] The strong-gravity regime associated with the initial
state leads to particle horizons, and spatial inhomogeneities are
constrained to have super-horizon scale as the initial singularity
is approached.

\item[(vi)] Increased strength of the gravitational field and the
collapse of particle horizons lead to asymptotic silence, and on
the scale of the particle horizon solutions therefore are
asymptotically SH.

\item[(vii)] The past attractor describing asymptotic spatially
inhomogeneous dynamics is thus given by the generalized Mixmaster
attractor.

\item[(viii)] Final singularities are in essence the time reverse
of initial singularities, and so we expect the above ideas to apply
there too, and conversely that ideas from gravitational collapse
can throw some light on cosmological initial singularities. In
particular, in generic gravitational collapse, angular momentum
plays a fundamental physical r\^{o}le, so this should also be true
in many time-reversed cases, i.e., at the initial singularity; but
when this is true, matter is dynamically important, in contrast to
the cases considered above.
\end{itemize}
\vspace{2mm}

Each of these issues needs to be investigated and given a precise
mathematical statement; e.g, the last may relate to the hypothesis
that the tilt transition sets can be interpreted as
physical/dynamical effects of vorticity in the matter fluid and
associated transverse peculiar velocity components. In each case we
wish to link our results and conjectures to physical ideas. The
challenge is to explain the difference between the past attractor
for vacuum $G_{0}$ models and the past attractor for perfect fluid
$G_{0}$ models [\,compare Eqs.~(\ref{e35}) and~(\ref{e141})\,] using
physical principles. It is often stated that ``matter (energy) does
not matter'' in the approach to the initial singularity --- this
view is reflected in our past asymptotic limits for the matter
variables $\Om$ and $\Oml$. But, perhaps --- and this is rather
heuristic and speculative at this stage ---, ``matter linear
momentum does matter'' and/or ``matter angular momentum does
matter''.

In the end the major physical statements are

\begin{itemize}
\item {\em Ultra-strong gravitational fields will occur in the
early Universe, associated with local restrictions on causality\/}.

\item {\em Propagating gravitational waves are not important in the
cosmological context, but tidal forces are, and indeed are often
more important than the gravitational fields caused directly by the
matter}.

\item {\em The relation between tidal forces and vorticity in the
 matter fluid is unclear and may contain some of the most
 interesting physics\/}.
\end{itemize}

The relation between them is that --- if our conjectures are
correct --- in the early Universe, energy and information mainly
propagate along timelike worldlines rather than along null
rays. When matter moves relative to the irrotational timelike
reference congruence, as must be the case when vorticity in the
matter fluid is dynamically important, then the energy and
information will flow with the matter. The primary effect of the
gravitational field is in determining the motion of the matter
through Coulomb-like effects; on the other hand, the effect of the
matter on the gravitational field is primarily through
concentrating that field into small regions, while conserving the
constraints which embody the Gau\ss\ law underlying the
Coulomb-like behavior. The effect of spatial curvature is to
generate oscillatory behavior in tidal forces as this
concentration takes place, as seems to be characteristic of generic
cosmological initial singularities; but this is not wavelike in the
sense of conveying information to different regions, it is just a
localized oscillation.

It is issues such as these that need to be investigated when
further developing the themes studied here.

\section*{Acknowledgments}
We thank Woei Chet Lim for many helpful discussions and for
performing numerical simulations, and Joshua Horwood for preparing
the figures. C.U. and J.W. appreciate numerous stimulating
discussions during February 2003 with participants of the workshop
on ``Mathematical Aspects of General Relativity'', which helped
clarify our understanding of $G_{0}$ cosmologies and contributed to
the final form of this paper. This workshop was held at the
Mathemati\-sches Forschungsinstitut, Oberwolfach, Germany, whose
generous hospitality is gratefully acknowledged.
H.v.E. acknowledges repeated kind hospitality by the Department of
Physics, University of Karlstad, Sweden. C.U. was in part supported
by the Swedish Science Council. J.W. was in part supported by a
grant from the Natural Sciences and Engineering Research Council of
Canada.

\appendix
\section{Appendix}

\subsection{Propagation of constraints}
\lb{subsec:consprop}

\noindent
{\em Propagation of dimensional constraints\/}:
\nopagebreak
\bea
\lb{ccomdot}
\vece_{0}[\,(C_{\rm com})_{\alpha\beta}(f)\,]
& = & (C_{\rm com})_{\alpha\beta}[\,\vece_{0}(f)\,]
- 2H\,(C_{\rm com})_{\alpha\beta}(f) + 2\sig_{[\alpha}{}^{\gam}\,
(C_{\rm com})_{\beta]\gam}(f) \nonumber\\
& & \hsp5 - \ 2\eps_{\gam\delta [\alpha}\,\Om^{\gam}\,
(C_{\rm com})_{\beta]}{}^{\delta}(f) \\
\lb{cgaussdot}
\vece_{0}(C_{\rm G})
& = & -\,2H\,(C_{\rm G}) + 2\,(\vece_{\alpha}+2\udot_{\alpha}
-2a_{\alpha})\,(C_{\rm C})^{\alpha} + \eps^{\alpha\beta\gam}\,
(C_{\rm com})_{\alpha\beta}(\Om_{\gam}) \\
\lb{ccodaccidot}
\vece_{0}(C_{\rm C})^{\alpha}
& = & -\,[\ 4H\,\d^{\alpha}{}_{\beta}
+ \sig^{\alpha}{}_{\beta}
- \eps^{\alpha}{}_{\gam\beta}\,\Om^{\gam}\ ]\,
(C_{\rm C})^{\beta}
- \sfrac{1}{6}\,(\d^{\alpha\beta}\,\vece_{\beta}
-2\udot^{\alpha})\,(C_{\rm G}) \nonumber\\
& & \hsp5 + \ \sfrac{1}{2}\,
\eps^{\alpha\beta\gam}\,(\vece_{\beta}+\udot_{\beta}
-3a_{\beta})\,(C_{\rm J})_{\gam}
- \sfrac{3}{2}\,n^{\alpha}{}_{\beta}\,
(C_{\rm J})^{\beta} + \sfrac{1}{2}\,n_{\beta}{}^{\beta}\,
(C_{\rm J})^{\alpha} \\
& & \hsp5 - \ (C_{\rm com})^{\alpha}{}_{\beta}
(\udot^{\beta}-a^{\beta})
- \sfrac{1}{2}\,\eps^{\alpha\beta\gam}\,
(C_{\rm com})_{\beta\delta}(n_{\gam}{}^{\delta})
+ \sfrac{1}{4}\,\eps^{\beta\gam\delta}\,
(C_{\rm com})_{\beta\gam}(n^{\alpha}{}_{\delta}) \nonumber \\
\lb{cjac1dot}
\vece_{0}(C_{\rm J})^{\alpha}
& = & -\,[\ 2H\,\d^{\alpha}{}_{\beta}
- \sig^{\alpha}{}_{\beta}
- \eps^{\alpha}{}_{\gam\beta}\,\Om^{\gam}\ ]\,
(C_{\rm J})^{\beta}
- \sfrac{1}{2}\,\eps^{\alpha\beta\gam}\,
(C_{\rm com})_{\beta\gam}(H) \nonumber\\
& & \hsp5  + \sfrac{1}{2}\,\eps^{\alpha\beta\gam}\,
(C_{\rm com})_{\beta\delta}(\sig_{\gam}{}^{\delta})
- \sfrac{1}{4}\,\eps^{\beta\gam\delta}\,
(C_{\rm com})_{\beta\gam}(\sig^{\alpha}{}_{\delta})
- (C_{\rm com})^{\alpha}{}_{\beta}(\Om^{\beta}) \ .
\eea

\noindent
{\em Propagation of dimensionless gauge fixing condition\/}:
\be
\lb{gfcpropsa}
\ptl_{t}({\cal C}_{\Udot})_{\alpha}^{\rm sv}
= -\,(\d_{\alpha}{}^{\beta} + \Sig_{\alpha}{}^{\beta}
- \eps_{\alpha\gam}{}^{\beta}\,R^{\gam})\,
({\cal C}_{\Udot})_{\beta}^{\rm sv} \ .
\ee
%

\subsection{Hubble-normalized relativistic Euler equations}
\lb{subsec:Euler}
Upon substitution of the matter variables (\ref{dlpqpi}),
Eqs.~(\ref{dlomdot}) and (\ref{dlvdot}) assume the explicit form
\bea
\lb{dlomdotf}
\parb_{0}\Om
& = & -\,\frac{\gam}{G_{+}}\,v^{\alpha}\,\parb_{\alpha}\Om
+ G_{+}^{-1}\,[\,2G_{+}q - (3\gam-2) - (2-\gam)\,v^{2}
- \gam\,(\Sig_{\alpha\beta}v^{\alpha}v^{\beta}) \nonumber \\
& & \hspace{35mm} - \ \gam\,(\parb_{\alpha}-2r_{\alpha}
+2\Udot_{\alpha}-2A_{\alpha})\,v^{\alpha}
+ \gam\,v^{\alpha}\,\parb_{\alpha}\ln G_{+}\,]\,\Om \\
\lb{dlvdotf}
\parb_{0}v^{\alpha}
& = & -\,v^{\beta}\,\parb_{\beta}v^{\alpha}
+ \d^{\alpha\beta}\,\parb_{\beta}\ln G_{+}
- \frac{(\gam-1)}{\gam}\,(1-v^{2})\,\d^{\alpha\beta}\,
(\parb_{\beta}\ln\Om-2r_{\beta}) \nonumber \\
& & + \ G_{-}^{-1}\,[\,(\gam-1)\,(1-v^{2})\,(\parb_{\beta}
v^{\beta}) - (2-\gam)\,v^{\beta}\,\parb_{\beta}\ln G_{+}
\nonumber \\
& & \hspace{15mm}
+ \ \frac{(\gam-1)}{\gam}\,(2-\gam)\,(1-v^{2})\,v^{\beta}\,
(\parb_{\beta}\ln\Om-2r_{\beta})
+ (3\gam-4)\,(1-v^{2}) \nonumber \\
& & \hspace{15mm} + \ (2-\gam)\,(\Sig_{\beta\gam}v^{\beta}
v^{\gam}) + G_{-}\,(\Udot_{\beta}v^{\beta})
+ [G_{+}-2(\gam-1)]\,(A_{\beta}v^{\beta})\,]\,v^{\alpha}
\nonumber \\
& & - \ \Sig^{\alpha}{}_{\beta}\,v^{\beta}
+ \eps^{\alpha}{}_{\beta\gam}\,R^{\beta}\,v^{\gam}
- \Udot^{\alpha} - v^{2}\,A^{\alpha}
+ \eps^{\alpha\beta\gam}\,N_{\beta\delta}\,v_{\gam}\,v^{\delta}
\ .
\eea
Using $v_{\alpha} = v\,e^{\alpha}$, $e_{\alpha}e^{\alpha} = 1$, we
easily obtain from Eq.~(\ref{dlvdotf})
\bea
\lb{vsqdot}
\parb_{0}v^{2}
& = & -\,v^{\alpha}\parb_{\alpha}v^{2}
 + \frac{2}{G_{-}}\,(1-v^{2})\,[\,v^{\alpha}\,
\parb_{\alpha}\ln G_{+} + (\gam-1)\,v^{2}\,
(\parb_{\alpha}v^{\alpha}) \nonumber \\
& & \hspace{25mm} - \ \frac{(\gam-1)}{\gam}\,(1-v^{2})\,
v^{\alpha}\,(\parb_{\alpha}\ln\Om-2r_{\alpha}) \nonumber \\
& & \hspace{25mm} + \ (3\gam-4)\,v^{2}
- (\Sig_{\alpha\beta}v^{\alpha}v^{\beta}) - G_{-}\,(\Udot_{\alpha}
v^{\alpha}) - 2\,(\gam-1)\,v^{2}\,(A_{\alpha}v^{\alpha})\,] \\
\lb{dledotf}
\parb_{0}e^{\alpha}
& = & -\,v\,e^{\beta}\,\parb_{\beta}e^{\alpha}
+ \frac{1}{v}\,p^{\alpha\beta}\,\parb_{\beta}\ln G_{+}
- \frac{1}{v}\,\frac{(\gam-1)}{\gam}\,(1-v^{2})\,p^{\alpha\beta}\,
(\parb_{\beta}\ln\Om-2r_{\beta}) \nonumber \\
& & \hsp5 - \ p^{\alpha}{}_{\beta}\,\Sig^{\beta}{}_{\gam}\,e^{\gam}
+ s^{\alpha}{}_{\beta}\,R^{\beta}
- \frac{1}{v}\,p^{\alpha}{}_{\beta}\,\Udot^{\beta}
- v\,p^{\alpha}{}_{\beta}\,A^{\beta}
+ v\,s^{\alpha}{}_{\beta}\,N^{\beta}{}_{\gam}\,e^{\gam} \ .
\eea
Here $p^{\alpha}{}_{\beta} := \d^{\alpha}{}_{\beta} -
e^{\alpha}e_{\beta}$ and $s^{\alpha}{}_{\beta} :=
\eps^{\alpha}{}_{\beta\gam}\,e^{\gam}$.

\subsection{Hubble-normalized curvature variables}
\lb{subsec:siweyl}
The Hubble-normalized 3-Ricci curvature of a spacelike 3-surface
${\cal S}$:$\{t=\mbox{constant}\}$ is defined through the
symmetric-tracefree and trace parts. The trace part $\ck$ was given
in Eq.~(\ref{omkdef}), while the tracefree part is given by
\be
\lb{dltrfr3ric}
{\cal S}_{\alpha\beta}
:=  -\,\sfrac{1}{3}\,\eps^{\gam\delta}{}_{\la\alpha}\,
(\parb_{|\gam|}-r_{|\gam|}-2A_{|\gam|})\,N_{\beta\ra\delta}
+ \sfrac{1}{3}\,(\parb_{\la\alpha}-r_{\la\alpha})\,A_{\beta\ra}
+ \sfrac{2}{3}\,N_{\la\alpha}{}^{\gam}\,N_{\beta\ra\gam}
- \sfrac{1}{3}\,N_{\gam}{}^{\gam}\,N_{\la\alpha\beta\ra}\ .
\ee
The quantities ${\cal S}_{\alpha\beta}$ and $\ck$ satisfy the
Hubble-normalized twice-contracted 3-Bianchi identity given by
\be
\lb{dl3bianid}
0 \equiv (\parb_{\beta}-2r_{\beta}-3A_{\beta})\,{\cal S}^{\alpha\beta}
- \eps^{\alpha\beta\gam}\,N_{\beta\delta}\,{\cal S}_{\gam}{}^{\delta}
+ \sfrac{1}{3}\,\d^{\alpha\beta}\,(\parb_{\beta}-2r_{\beta})\,\ck \ .
\ee
Employing Eq.~(\ref{dltrfr3ric}), we can write the evolution equation
for $\Sig_{\alpha\beta}$ in the alternative form
\bea
\lb{dlsigdot2}
\parb_{0}\Sig^{\alpha\beta}
& = & (q-2)\,\Sig^{\alpha\beta} - 3\,({\cal S}^{\alpha\beta}
-\Pi^{\alpha\beta}) + \eps^{\gam\delta\la\alpha}\,[\,2R_{\gam}\,
\Sig^{\beta\ra}{}_{\delta} - N^{\beta\ra}{}_{\gam}\,
\Udot_{\delta}\,] \nonumber \\
& & \hsp5 + \ (\d^{\gam\la\alpha}\,\parb_{\gam}-r^{\la\alpha}
+\Udot^{\la\alpha}+A^{\la\alpha})\,\Udot^{\beta\ra} \ .
\eea
The conformal curvature properties of a spacelike 3-surface ${\cal
S}$:$\{t=\mbox{constant}\}$ are encoded in the Hubble-normalized
3-Cotton--York tensor
\be
\lb{dl3ct}
{\cal C}_{\alpha\beta} := \eps^{\gam\delta}{}_{\la\alpha}\,
(\parb_{|\gam|}-2r_{|\gam|}-A_{|\gam|})\,{\cal S}_{\beta\ra\delta}
- 3N_{\la\alpha}{}^{\gam}\,{\cal S}_{\beta\ra\gam}
+ \sfrac{1}{2}\,N_{\gam}{}^{\gam}\,{\cal S}_{\alpha\beta} \ .
\ee
The Hubble-normalized Weyl curvature variables take the explicit
form
\bea
\lb{dlele}
{\cal E}_{\alpha\beta}
& = & {\cal S}_{\alpha\beta} + \sfrac{1}{3}\,\Sig_{\alpha\beta}
- \sfrac{1}{3}\,\Sig_{\la\alpha}{}^{\gam}\,
\Sig_{\beta\ra\gam} - \sfrac{1}{2}\,\Pi_{\alpha\beta} \\
\lb{dlmag}
{\cal H}_{\alpha\beta}
& = & \sfrac{1}{3}\,\eps^{\gam\delta}{}_{\la\alpha}\,
(\parb_{|\gam|}-r_{|\gam|}-A_{|\gam|})\,\Sig_{\beta\ra\delta}
- N_{\la\alpha}{}^{\gam}\,\Sig_{\beta\ra\gam}
+ \sfrac{1}{6}\,N_{\gam}{}^{\gam}\,\Sig_{\alpha\beta} \ ,
\eea
with ${\cal S}_{\alpha\beta}$ defined in Eq.~(\ref{dltrfr3ric}).

\subsection{Curvature transitions}
\lb{subsec:curvtr}
Although the relation (\ref{curvtr}) implicitly gives the rule for
the relationship between two Kasner epochs, $k$ is not particularly
suitable for explicitly describing the ``Kasner transformation
law'' for curvature transitions. However, that law can be elegantly
obtained in the present dynamical systems framework as follows.
The solutions on the $\cT_{N_1}$-subset are determined by
$$
\sfrac{1}{12}\,N_{1}^{2} = 1 - \Sig^{2} \ ,
$$
and
$$
\Sig_{1} = -\,4 + 3Z \ , \hsp5
\Sig_{2} = 2 - \sfrac{3}{2}\,r_{+}\,Z \ , \hsp5
\Sig_{3} = 2 - \sfrac{3}{2}\,r_{-}\,Z \ , \hsp5
\ptl_{t}Z = -\,2(1-\Sig^{2})\,Z \ ,
$$
where $r_{\pm} := 1\pm\sqrt{1-\alpha^{2}}$, $\alpha \in [0,1]$, is
a constant, and where $1-\Sig^{2} =
\sfrac{3}{4}\,[\,(\alpha-2)Z+2\,]\,[\,(\alpha+2)Z-2\,]$. An orbit
starts (with time direction reversed toward the past) from a Kasner
point where $Z = Z_{-} = 2/(2+\alpha)$, and ends at a Kasner point
where $Z = Z_{+} = 2/(2-\alpha)$. That is, $Z$ is a parameter on
the individual orbits that increases monotonically from $Z_{-}$ to
$Z_{+}$ toward the past, while $\alpha$ labels the different
orbits. The value $Z = 0$ determines the point $P_{1}$ outside the
Kasner circle in Fig.~\ref{fig:kasnc}. It is possible to express the
constant $\alpha$, and, hence, $r_{\pm}$, in terms of the standard
Kasner parameter $u \geq 1$ (see, e.g., BKL~\ct{bkl70}, p.~528),
where we assume that we are considering orbits that originate from
sector $(123)$ and thus that $p_{1} < p_{2} < p_{3}$. Then $(r_{+},
\alpha, r_{-}) = 2(u^2,u,1)/(1+u^{2})$, and thus $\alpha$ might be
viewed as a compactified Kasner parameter. The above formulae
directly yield the standard transformation laws from the initial
Kasner exponents $\{\,p_{\alpha}\,\}$ to the final Kasner exponents
$\{\,p_{\alpha}^{\prime}\,\}$ for orbits originating from sector
$(123)$:
\be
p_{1}^{\prime} = \frac{-p_{1}}{1+2p_{1}} \ , \hsp5
p_{2}^{\prime} = \frac{p_{2}+2p_{1}}{1+2p_{1}} \ , \hsp5
p_{3}^{\prime} = \frac{p_{3}+2p_{1}}{1+2p_{1}} \ .
\ee
Curvature transitions from other sectors than $(123)$ are easily
obtained through permutations.

\subsection{Inverting the commutator equations}
\lb{subsec:comsolve}
The commutator equations can be used to solve for the connection
variables in terms of the frame variables. Let us introduce the
{\em dual frame variables\/} $e^{\alpha}{}_{i}$ of
$e_{\alpha}{}^{i}$, and their Hubble-normalized counterparts
$E^{\alpha}{}_{i} := H\,e^{\alpha}{}_{i}$. The dual frame variables
can be used to conveniently describe the line element, giving
\be
\lb{ds2gen}
{\rm d}s^{2} = H^{-2}\left[\,-\,\cn^{2}\,{\rm d}t^{2}
+ \d_{\alpha\beta}\,E^{\alpha}{}_{i}\,E^{\beta}{}_{j}\,
(N^{i}{\rm d}t+{\rm d}x^{i})\,(N^{j}{\rm d}t+{\rm d}x^{j})
\,\right] \ .
\ee
As $E_{\alpha}{}^{i}$ and $E^{\alpha}{}_{i}$ satisfy the relations
$E^{\alpha}{}_{i}\,E_{\beta}{}^{i} = \d^{\alpha}{}_{\beta}$ and
$E^{\alpha}{}_{i}\,E_{\alpha}{}^{j} = \d_{i}{}^{j}$, one finds with
Eq.~(\ref{dl13comts}) that the $E^{\alpha}{}_{i}$ evolve according to
\be
\lb{dlinvfrmdot}
\parb_{0}E^{\alpha}{}_{i}
= -\,(q\,\d^{\alpha}{}_{\beta} - \Sig^{\alpha}{}_{\beta}
- \eps^{\alpha}{}_{\gam\beta}\,R^{\gam})\,E^{\beta}{}_{i}
+ \cn^{-1}\,E^{\alpha}{}_{j}\,\ptl_{i}N^{j} \ .
\ee
On the other hand, from appropriately inverting
Eqs.~(\ref{dl13comts}), (\ref{dl13comss}) and (\ref{dl13comtt31}) we
obtain the explicit expressions
\bea
q & = & \sfrac{1}{3}\,E^{\alpha}{}_{i}\,\parb_{0}E_{\alpha}{}^{i}
+ \sfrac{1}{3}\,\cn^{-1}\,\ptl_{i}N^{i} \\
\Sig_{\alpha\beta}
& = & -\,\d_{\gam\la\alpha}\,E^{\gam}{}_{i}\,
\parb_{0}E_{\beta\ra}{}^{i} 
- \cn^{-1}\,\d_{\gam\la\alpha}\,E^{\gam}{}_{i}\,
\parb_{\beta\ra}N^{i} \\
R^{\alpha}
& = & \sfrac{1}{2}\,\eps^{\alpha}{}_{\beta}{}^{\gam}\,
E^{\beta}{}_{i}\,\parb_{0}E_{\gam}{}^{i}
+ \sfrac{1}{2}\,\cn^{-1}\,\eps^{\alpha}{}_{\beta}{}^{\gam}\,
E^{\beta}{}_{i}\,\parb_{\gam}N^{i} \\
A_{\alpha} + r_{\alpha}
& = & \sfrac{1}{2}\,E^{\beta}{}_{i}\,\parb_{\alpha}E_{\beta}{}^{i}
- \sfrac{1}{2}\,\ptl_{i}E_{\alpha}{}^{i} \\
N^{\alpha\beta}
& = & E^{(\alpha}{}_{i}\,\epsilon^{\beta)\gam\delta}\,
\parb_{\gam}E_{\delta}{}^{i} \lb{e165}\\
\Udot_{\alpha} - r_{\alpha}
& = & \parb_{\alpha}\ln\cn \ .
\eea
When working in the {\em separable volume gauge\/}, determined by
conditions~(\ref{N1}) and (\ref{zeroshift}), the line element takes the
form
\be
\lb{ds2}
{\rm d}s^{2} = H^{-2}\left[\,-\,{\rm d}t^{2}
+ \d_{\alpha\beta}\,E^{\alpha}{}_{i}\,E^{\beta}{}_{j}\,
{\rm d}x^{i}\,{\rm d}x^{j}\,\right] \ .
\ee
Combining the above result for $A_{\alpha} + r_{\alpha}$ with the
constraint~(\ref{dlsepcon}) makes it then possible to solve for
$r_{\alpha}$ and $A_{\alpha}$ separately:
\bea
r_{\alpha} & = & \sfrac{1}{3}\,E^{\beta}{}_{i}\,E_{\alpha}{}^{j}
\,\ptl_{j}E_{\beta}{}^{i}
+ \sfrac{1}{3}\,E_{\alpha}{}^{i}\,\ptl_{i}\ln\hat{m}  \\
A_{\alpha} & = & \sfrac{1}{6}\,E^{\beta}{}_{i}\,E_{\alpha}{}^{j}
\,\ptl_{j}E_{\beta}{}^{i}
- \sfrac{1}{2}\,\ptl_{i}E_{\alpha}{}^{i}
- \sfrac{1}{3}\,E_{\alpha}{}^{i}\,\ptl_{i}\ln\hat{m}  \ .
\eea
%



\end{document}